\documentclass[12pt]{article}
\newcommand{\Q}{\ensuremath{\mathcal{Q}}}
\newcommand{\Qt}{\ensuremath{\tilde{\mathcal{Q}}}}
\newcommand{\tn}{\ensuremath{t_{0}}}
\newcommand{\xtn}{\ensuremath{x(t_{0})}}
\newcommand{\xn}{\ensuremath{x_{0}}}
\newcommand{\xij}{\ensuremath{x_{ij}}}
\newcommand{\li}{\ensuremath{\lambda_{i}}}
\newcommand{\ri}{\ensuremath{\rho_{i}}}
\newcommand{\lijk}{\ensuremath{\lambda_{ijk}}}
\newcommand{\rijk}{\ensuremath{\rho_{ijk}}}
\newcommand{\xiij}{\ensuremath{\xi_{ij}}}
\newcommand{\xiijk}{\ensuremath{\xi_{ijk}}}
\newcommand{\Rij}{\ensuremath{R_{ij}}}
\newcommand{\Rtwo}{\ensuremath{R_{1234}}}
\usepackage{epsfig}
\begin{document}
\begin{titlepage}
\vspace{2cm}
\title{
\vskip -50pt 
\begin{flushright}
\begin{normalsize}
 \ DAMTP--2000--8,
{\tt hep-th/0001155}\\ 
\end{normalsize}
\end{flushright} 
\vskip 40pt
\begin{bfseries}
Classical Supersymmetric Mechanics
\end{bfseries}
}
\vspace{2cm}
\author{\\ R. Heumann \footnote{email: R. Heumann@damtp.cam.ac.uk}\quad 
and\quad N.S. Manton \footnote{email: N.S.Manton@damtp.cam.ac.uk}
\bigskip\\
\textit{Department of Applied Mathematics and Theoretical Physics}\\ 
\textit{University of Cambridge}\\
\textit{Wilberforce Road, Cambridge CB3 0WA, England}\\
}
\date{\large January 21, 2000}
\maketitle
\thispagestyle{empty}
\begin{abstract}
\noindent
We analyse a supersymmetric mechanical model derived from
$(1\!+\!1)$-dimensional field theory with Yukawa interaction, assuming
that all physical variables take their values in a Grassmann algebra
$\mathcal{B}$. Utilizing the symmetries of the model we demonstrate
how for a certain class of potentials the equations of motion can be
solved completely for any $\mathcal{B}$. In a second approach we
suppose that the Grassmann algebra is finitely generated, decompose
the dynamical variables into real components and devise a
layer-by-layer strategy to solve the equations of motion for arbitrary
potential. We examine the possible types of motion for both bosonic
and fermionic quantities and show how symmetries relate the former to
the latter in a geometrical way. In particular, we investigate
oscillatory motion, applying results of Floquet theory, in order to
elucidate the role that energy  variations of the lower order quantities 
play in determining the quantities of higher order in $\mathcal{B}$.
\end{abstract}
\end{titlepage}
\section{Introduction}
Classical supersymmetry sets out to extend the unified treatment of
bosonic and fermionic quantities in the usual QFT framework to the
classical level. Normally, in semiclassical treatments the fermionic
variables are set to zero as soon as the supersymmetric theory has
been constructed. The usual argument goes that since we cannot find
classical fermions in nature, fermionic quantities should be omitted
altogether at the classical level. 

However, this is far from necessary. In fact, a consistent
approach to classical supersymmetry has long been available -- for a
review see e.g. the book by de Witt \cite{deWitt}. Fermionic
quantities are then treated as anticommuting variables taking values
in a Grassmann algebra $\mathcal{B}$. Grassmann-valued mechanics has
been analysed in the works of Berezin and Marinov \cite{Berezin} and
Casalbuoni \cite{Casalbuoni} and later by Junker and Matthiesen
\cite{Junker}. A main difference to our work is that both
\cite{Berezin} and \cite{Casalbuoni} do not distinguish clearly 
between generators of the algebra and dynamical quantities and thus
define the Grassmann algebra $\mathcal{B}$ rather implicitly. The fact
that the bosonic variables take values in the even part of the same
algebra $\mathcal{B}$ is not apparent in these works, although both
recognize that the bosonic variables cannot be real functions anymore --
without, however, elaborating on this fact. A central aim of this
paper is therefore to make sense of the \textit{general}
Grassmann-valued equations of motion, including the fermionic ones,
and to find ways to their solution, which is done in \cite{Berezin} and
\cite{Casalbuoni} only in very special cases. Junker and Matthiesen,
who investigate a similar mechanical model, achieve a more general
solution than in \cite{Berezin} and \cite{Casalbuoni}, but again under the
(implicit) assumption that the Grassmann algebra is spanned by only
two generators identified with the fermionic dynamical variables. We
can confirm most of their results (in different form, though, due to a
different choice of variables) as special cases of our solutions. 
However, we disagree about some details, in particular, concerning the
case of zero energy. 

\bigskip
The mechanical model that we study here is the supersymmetric motion
of a particle in a one-dimensional potential, derived by dimensional
reduction from the usual $N=2$ supersymmetric $(1+1)$-dimensional
field theory with Yukawa interaction. A slightly different version of
this model was investigated in \cite{Manton}, where a different
concept of reality was used that led to a negative potential in the
Lagrangian. The approach taken here stays closer to the usual case
with the positive potential. 

An important result of \cite{Manton} was that a complete solution for
the particle motion could be found on the assumption that the
underlying algebra $\mathcal{B}$ has only two generators. This led
to relatively simple results, however is unnecessarily restrictive.

Here we show first that for a large class of potentials the solution
to the equations of motion can be found for any $\mathcal{B}$ and
depends only on a small number of $\mathcal{B}$-valued constants of
integration, one of which is a Grassmann energy $E$. 

To deal with essentially arbitrary potentials we adopt a second method
which is closer to that of \cite{Manton}, although we need not restrict
ourselves to two generators: Choosing the Grassmann algebra to be
finitely generated, with $n$ generators, we split all dynamical
quantities and equations into their real components, named according
to the number of generators involved in the corresponding monomial. 
Then, beginning from the zeroth order equation, which can be seen as a
form of Newton's equation, we subsequently work our way up to higher
and higher orders, utilizing the solutions already found for the lower
levels. This layer-by-layer strategy allows us to solve the equations
of motion for any potential with reasonable mathematical properties.
 
The existence of a complete solution to the coupled system
of equations of motion looks surprising in view of the increasingly
large number of equations involved for large $n$. However, on second
thoughts it is not so unexpected: Due to our first solution method we
know that a full Grassmann solution can be found in many cases, the
decomposition of which should give us exactly the component solutions
obtained by the second method -- which it does indeed as we shall
demonstrate. 
 
A final word has to be said about the assumption of only a finite number
of generators since it has been claimed that this must necessarily
lead to contradictions: Emphasizing that our paper deals with the
\textit{classical} theory we do not find this to be true. 

\bigskip
We begin our analysis in section \ref{susymechanics} by presenting the
Lagrangian and the equations of motion that we will be concerned with
in this paper. Essential for solving these equations are the
symmetries and associated Noether charges of the Lagrangian which we
therefore examine in section \ref{grassmannsymmetries}. For a certain
class of potential functions, namely those for which a particular
integral can be calculated analytically, we describe in section
\ref{solfixpot} how the equations of motion can be solved completely
and illustrate this method for two exemplary potentials, the harmonic
potential $U(x)=kx$ and the hyperbolic potential $U(x)=c\tanh kx$.

Sections \ref{finite algebra} and \ref{solarbpot} are devoted to the
description of our layer-by-layer method which is then explicitly carried
out up to fourth order, and illustrated by the harmonic oscillator
case in section \ref{harmpot}.

Next, we investigate the symmetries in component form in section
\ref{symmetries}: While all component charges can be simply derived by
decomposing the original charges, they also reflect a huge number of
symmetries of the highest order component Lagrangian by which they can
be found using Noether's procedure. In addition to the symmetries
known from the original Lagrangian there appear extra classes of
symmetries which cannot be related to the former. It is possible to
trace them back to invariance under certain changes of base in the
Grassmann algebra, viewed as a $2^{n}$-dimensional vector space.

One question that immediately arises from the decomposition of the
physical quantities into components is addressed in section \ref{interpret}: 
How must we interpret the numerous new functions that arise from this
procedure? Utilizing the new symmetries found in section \ref{symmetries} 
we offer a geometric interpretation for the lowest order bosonic and
fermionic functions and describe the three basic types of motion
possible for them. The higher order quantities are interpreted here
from a different point of view: We see them as variations of the lower
order quantities with respect to the integration constants
involved in these functions, namely an initial time $t_{0}$ and
the \mbox{energy $E_{0}$.}

In section \ref{specialtypesofmotion} we apply our results to study
general oscillatory motion, by which we mean that the lowest order bosonic
function is periodic. The characteristic appearance of linear-periodic
terms is explained from both a physical and mathematical perspective
employing results of Floquet theory. Again, we choose a
particular potential to illustrate this in section \ref{hyperbolpot},
which allows us also to show that the two solution methods presented
in this paper coincide. 

A common restriction on both methods is that the energy $E_{0}$ of the
system must be positive, therefore the zero energy case has to be
discussed separately. We do this in section \ref{zeroenergy}.

Some ideas for further generalization and analysis conclude this paper.

\section{Supersymmetric Mechanics}
\label{susymechanics}
We start our discussion with the standard Lagrangian density for 
$(1+1)$-dimensional supersymmetric field theory with Yukawa
interaction:
\begin{displaymath}
\mathcal{L}=\frac{1}{2}\partial_{+}\phi\,\partial_{-}\phi - 
\frac{1}{2}U(\phi)^{2}+\frac{i}{2}\psi_{+}\,\partial_{-}\psi_{+} +
\frac{i}{2}\psi_{-}\,\partial_{+}\psi_{-} +
i\frac{dU}{d\phi}\psi_{+}\psi_{-},
\end{displaymath}
containing a real bosonic scalar field $\phi$, a real two-component fermionic
spinor field $\psi$ and a potential function $U(\phi)$;
$\partial_{\pm}$ are the light cone derivatives
$\partial_{t}\pm\partial_{x}$.

We assume that $\phi$ and the two components of $\psi$ take their
values in the real even and odd part, respectively, of an arbitrary
Grassmann algebra $\mathcal{B}$, and that the potential function
$U(\phi)$ can be expanded into a power series in $\phi$ with real
coefficients. Complex conjugation is defined such that
$(z_{1}\,z_{2})^{*}=z_{2}^{*}\,z_{1}^{*}$ in accordance with the
conventions in \cite{deWitt} so that the Lagrangian density is a real
function despite the $i$-factors that occur in front of the fermion terms.
As we are dealing with the classical case we assume further that all
fields and their derivatives commute or anticommute, depending in the
usual way on the bosonic or fermionic nature of the fields.

\bigskip
In this paper we shall be interested only in spatially independent
fields, i.e. 
$\partial_{x}\phi=\partial_{x}\psi_{+}=\partial_{x}\psi_{-}=0$.
As the fields are then functions of time only this leads us directly
from field theory to mechanics. It is therefore sensible to think of
the bosonic field $\phi$ as describing the one-dimensional motion of a
particle in the potential $U^2$. To support this notion we
will change the variable $\phi$ to $x$ for the rest of this paper.
We then obtain the following Lagrange function:
\begin{equation}
L=\frac{1}{2}\dot{x}^{2}-\frac{1}{2}U(x)^{2}+\frac{i}{2}\psi_{+}\dot{\psi}_{+}
+\frac{i}{2}\psi_{-}\dot{\psi}_{-}+i\,U'(x)\psi_{+}\psi_{-}.
\label{Lagrangian}
\end{equation}
where the dot denotes a time derivative and $U'$ means the derivative of
$U$ with respect to $x$.
Performing a formal variation of this Lagrangian with respect to the
variables $x$, $\psi_{+}$ and $\psi_{-}$ and neglecting total time
derivatives we derive the equations of motion for the system:
\begin{eqnarray}
\ddot{x}&=&-U(x)U'(x)+iU''(x)\psi_{+}\psi_{-}\label{x-equation}\\
\dot{\psi}_{+}&=&-U'(x)\psi_{-}\label{fermion-equation 1}\\
\dot{\psi}_{-}&=&\;\;\,U'(x)\psi_{+}.\label{fermion-equation 2}
\label{eqns of motion}
\end{eqnarray}
There is a slight ambiguity in these equations if the Grassmann
algebra is finitely generated and has an odd number of generators.
Then the two equations for the fermion variables will be 
determined only up to an arbitrary function of highest order in the
Grassmann algebra. We can think of this as a gauge degree of
freedom and will come back to this point later.

In order to solve the equations it is advisable to understand
the symmetries of the system first. This we will do in the next section.

\section{The symmetries of the model}
\label{grassmannsymmetries}
The first thing to notice is that the Lagrangian has no
explicit time dependence. We therefore have invariance under time
translation, leading to a conserved Hamiltonian as the corresponding
Noether charge. This we calculate to be
\begin{equation}
H=\frac{1}{2}\dot{x}^{2}+\frac{1}{2}U(x)^{2}-iU'(x)\psi_{+}\psi_{-};
\label{Hamiltonian}
\end{equation}
its conserved value we call the energy, denoted by $E$.
Note, however, that this Hamiltonian is an even Grassmann-valued
function and that therefore the conserved energy $E$ is an even
element of $\mathcal{B}$.

The operator $\frac{d}{dt}$ generates time translations, and acts on the
dynamical variables in the obvious way:
\begin{equation}
\label{timetranslation}
\Delta x = \eta\dot{x}, \: \Delta \psi_{1} = \eta\dot{\psi}_{1}, \:
\Delta \psi_{2} = \eta\dot{\psi}_{2},
\end{equation}
where $\eta$ is an infinitesimal even Grassmann parameter.

\bigskip
In addition to time translation invariance there are two further
independent symmetries of the Lagrangian, relating fermions and
bosons. These can be written in infinitesimal form as:
\begin{equation}
\begin{array}{rclrclrcl}
\delta x &=& i \epsilon \psi_{+},& \delta \psi_{+}&=& -\epsilon \dot{x},&
\delta \psi_{-}&=&-\epsilon U(x),\\
\tilde{\delta} x &=& i \epsilon \psi_{-},&\tilde{\delta} \psi_{+}&=& 
\epsilon U(x),&\tilde{\delta} \psi_{-}&=&-\epsilon \dot{x},
\end{array}
\label{SUSY transformation}
\end{equation}
where $\epsilon$ is an arbitrary infinitesimal odd Grassmann
parameter. These transformations lead only to a change in the
Lagrangian by a total time derivative. We can therefore apply Noether's
procedure and find the following charges:
\begin{eqnarray}
Q&=&\dot{x}\psi_{+} + U(x)\psi_{-}\label{q},\\
\tilde{Q}&=&\dot{x}\psi_{-} - U(x)\psi_{+},\label{qt}
\end{eqnarray}
the conservation of which can be easily shown using the equations of
motion. Note that the charges are odd elements of the Grassmann
algebra.

As for time translation invariance we now define two operators \Q\
and \Qt\ generating the two symmetry
transformations. From (\ref{SUSY transformation})  we can read off the
action of these operators on the dynamical variables:
\begin{equation}
\begin{array}{rclrclrcl}
\Q\ x &=& \psi_{+},&\Q\ \psi_{+}&=&\;\;\;i \dot{x},&\Q\ \psi_{-}&=&i U(x)\\
\Qt\ x &=& \psi_{-},&\Qt\ \psi_{+}&=&-i U(x),&\Qt\ \psi_{-}&=&i \dot{x}.
\end{array}
\label{SUSY representation}
\end{equation}

Using the action of the operators on $x$, $\psi_{+}$ and $\psi_{-}$ we
find that \Q, \Qt\ and $\frac{d}{dt}$ form a closed algebra
with the relations
\begin{displaymath}
\Q^{2}=i \frac{d}{dt},\; \Qt^{2}=i \frac{d}{dt},\; \{\Q,\Qt\}=0
\end{displaymath}
as long as the equations of motion are satisfied. Notice that
$\frac{d}{dt}$ commutes with everything.

This is the usual $N=2$ supersymmetry algebra in $0$ space dimensions,
and with (\ref{SUSY representation}) we have an \textit{\mbox{on-shell}}
representation of this algebra. Thus we will from now on speak of
supersymmetry transformations and call the associated charges
supercharges.

\bigskip
Beside time-translation invariance and the two supersymmetries there
exists still a further invariance, this time for the
fermionic functions only. In infinitesimal form it is given by the
transformation
\begin{equation}
\label{fermion rotation}
\tilde{\Delta} \psi_{+}=\eta \psi_{-},\; \tilde{\Delta}
\psi_{-}=-\eta \psi_{+}, 
\end{equation}
where $\eta$ again denotes an infinitesimal even Grassmann parameter. 
We can think of this as an internal rotation of the fermionic variables.
This invariance leads to a further conserved Noether charge:
\begin{equation}
R=i\psi_{+}\,\psi_{-}.\label{R}
\end{equation}
Comparison with (\ref{Lagrangian}) shows that the fermionic functions enter
the interaction term in the Lagrangian only via $R$. This leads
directly to the fact that the $x$-motion depends on the
fermionic functions only through this one constant. 

In addition, the Hamiltonian (\ref{Hamiltonian}) reveals that the only fermion
contribution to the total energy $E$ is through $R$. This is an
important simplification, and it will allow us to solve the equations of
motion completely for a number of potential functions $U$ without
restriction on the nature of the Grassmann algebra~$\mathcal{B}$.

\section{General solutions via Grassmann integrals}
\label{solfixpot}
We have already mentioned that the $x$-equation of motion (\ref{x-equation})
nearly decouples from the other two equations because the fermion
functions in the coupling term~$i\,U'(x)\,\psi_{+}\psi_{-}$ form a
Grassmann constant. It is therefore sensible to begin with this equation.

From the conserved Hamiltonian we know that
\begin{equation}
\dot{x}^{2}=2E-U(x)^{2}+2RU'(x).
\label{energy}
\end{equation}
For the next step we have to use that every Grassmann number $z$ can be
split into two parts, its 'body' and its 'soul': $z=z_{b}+z_{s}$.
The body is just the real number content of $z$, the soul is the remaining
linear combination of products of (odd) Grassmann generators and will be 
nilpotent if the Grassmann algebra is finitely generated. Note, however, 
that we do not need such a restriction for body and soul to be well-defined. 
A square root for a Grassmann quantity can be defined by its power series as
long as the body is positive. Since $R$ is the product of
two odd Grassmann terms, its body and therefore the body of the whole
third term on the right hand side of (\ref{energy}) is zero, leaving
the restriction $2E_{b}-U_{b}^{2}(x)> 0$ if $\dot{x}$ is to be
well-defined. This just means that the kinetic energy of the classical
particle moving in the potential $U^{2}$ has to be positive.

The resulting first order differential equation for $x$ is
\begin{displaymath}
\frac{dx}{dt}=\pm\sqrt{2E-U(x)^{2}+2RU'(x)}.
\end{displaymath}
Provided $\mathcal{B}$ is finitely generated, we may regard $x$ as
lying in the vector space $\mathcal{B}$ and apply the standard theory
of systems of ODE's to show that this equation has a unique solution
for any given initial data $x(\tn)$. 

The equation can formally be solved by separating variables: 
\begin{equation}
t-t_{0}=\pm\int_{x(t_{0})}^{x(t)} \frac{dx}{\sqrt{2E-U(x)^{2}+2RU'(x)}}.
\label{Grassmann integral}
\end{equation}
Note that while the left-hand side is just a real expression, the
right-hand side is a Grassmann integral of a Grassmann
integrand. Such an integral is defined as a line integral in the
Grassmann algebra, thought of as a finite or infinite-dimensional
vector space spanned by products of generators. As is shown in
\cite{deWitt} such an integral is independent of the actual path
relating start- and endpoint of the integral since the integrand is
Grassmann-even.

Now the integrand comes from an ordinary real function extended into
the full Grassmann algebra using its power series. We can therefore
sensibly ask whether there is an indefinite integral $F(x)$ to this
\textit{real} function. If so, we can extend \textit{this} function
back into the Grassmann algebra thus gaining an indefinite integral
for the Grassmann integrand. Because the integrand is even it then follows
\cite{deWitt} that the integral is given by the difference of $F$
evaluated at the start- and endpoints of the path:
\begin{displaymath}
t-t_{0}=\pm\left( F(x(t))-F(x(t_{0}))\right)
\end{displaymath}
If the function $F$ has an inverse, we finally get $x$ as a function
of $t$ and therefore the solution of (\ref{x-equation}). We will
illustrate this method below for two potential functions $U$.

\bigskip
The solution to the fermion equations (\ref{fermion-equation 1}) and
(\ref{fermion-equation 2}) is now easy. From the solution for $x(t)$
we can immediately calculate $\dot{x}(t)$ and $U(x(t))$. Using the
explicit formulae (\ref{q}) and (\ref{qt}) for the two conserved
supercharges $Q$ and $\tilde{Q}$ we can evaluate the linear combinations
\begin{eqnarray*}
Q\dot{x}-\tilde{Q}\,U(x)&=&2H\psi_{+},\\
Q\,U(x)+\tilde{Q}\dot{x}&=&2H\psi_{-},
\end{eqnarray*}
where we have used that $R\,\psi_{+}=R\,\psi_{-}=0$ which follows from
(\ref{R}). Since the Hamiltonian $H$ has the constant value $E$ we
deduce that 
\begin{eqnarray}
\psi_{+}&=&\frac{Q\dot{x}-\tilde{Q}\,U(x)}{2E},\label{fermion1}\\
\psi_{-}&=&\frac{Q\,U(x)+\tilde{Q}\dot{x}}{2E}.\label{fermion2}
\end{eqnarray}
This effectively solves (\ref{fermion-equation 1}) and
(\ref{fermion-equation 2}), as the values of $Q$ and $\tilde{Q}$ are
determined by the data at the initial time $t_{0}$. There is one
subtle point here, however, since the division by $2E$ is only
possible as long as $E_{b}$, the body of $E$, which can be interpreted
as the classical energy of the particle, is non-zero. Indeed, when we
analyse the solution for finitely generated Grassmann algebras later,
we will find again that the $(E_{b}\!=\!0)$-case is special.

A second point which has to be mentioned here is that we have treated
$R$ as an independent parameter. However, $R$ is \textit{determined} by the two
fermionic quantities $\psi_{+}$ and $\psi_{-}$ and, in effect, by the
two supercharges $Q$ and $\tilde{Q}$:
\begin{displaymath}
R=i\frac{Q\tilde{Q}}{2E},
\end{displaymath}
which can be verified using the definition of $R$ in (\ref{R}). This
does not render our solution invalid but it shows that $R$ is not a
parameter that can be chosen independently. But let us now turn to
some examples:

\subsection{The harmonic potential $U(x)=kx$}
\label{harmonicpotential}
We start with one of the easiest problems, the harmonic oscillator with
potential \mbox{$U(x)=kx$,} \mbox{$k$ being} real. Here the integrand in 
(\ref{Grassmann integral}) is given by:
\begin{displaymath}
\frac{1}{\sqrt{(2E+2kR)-(kx)^2}},
\end{displaymath}
where $E$ is the constant Grassmann energy. Note that due to the
special nature of the potential we can combine the $U'(x)$-term $2kR$
with the energy~$E$ into one overall constant. If we treat the integrand
as an ordinary real function it has an indefinite integral:
\begin{displaymath}
\int \frac{1}{\sqrt{a^2-b^2x^2}}\, dx = \frac{1}{b} \arcsin
\left(\frac{b}{a}x\right),
\end{displaymath}
so that we can write
\begin{displaymath}
t-t_{0}=\pm \left[ \frac{1}{k} \arcsin (\frac{k}{\sqrt{2E+2kR}}\, x)
\right]^{x(t)}_{x(t_{0})}.
\end{displaymath}
This formula can be inverted to yield the solution
\begin{equation}
x(t)=\xtn \cos k(t-\tn) \pm \frac{v(\tn)}{k} \sin k(t-\tn),\label{hx-solution}
\end{equation}
where we have denoted the constant $\sqrt{2E+2kR-k^2\xtn^2}$ by  $v(\tn)$.
This looks formally like the usual harmonic oscillator solution, and
it is periodic with period $\frac{2\pi}{k}$, but all
terms are Grassmann-valued and not just real functions.

\bigskip
For the two fermion terms we need only calculate $\dot{x}$ and
$U(x)$ and insert them into equations (\ref{fermion1}) and
(\ref{fermion2}), which leaves us with:
\begin{eqnarray}
\psi_{+}&=&p_{+}\cos k(t-\tn) -p_{-}\sin k(t-\tn)\label{hpsi1-solution}, \\
\psi_{-}&=&p_{-}\cos k(t-\tn) +p_{+}\sin k(t-\tn)\label{hpsi2-solution},
\end{eqnarray}
where
\begin{displaymath}
p_{+}=\frac{\pm Q v(\tn)-\tilde{Q}k\xtn}{2E},\;
p_{-}=\frac{\pm \tilde{Q} v(\tn)+Q k\xtn}{2E}.
\end{displaymath}
With (\ref{hx-solution}), (\ref{hpsi1-solution}) and
(\ref{hpsi2-solution}) we have found the complete solution to the
equations of motion for $E_{b}>0$, independent of the nature of the Grassmann
algebra $\mathcal{B}$.

\subsection{The hyperbolic potential $U(x)=c \tanh kx$}
\label{grassmannhyperbol}
As a second example we choose the hyperbolic potential $U(x)=c \tanh
kx$, with $c$ and $k$ both real. The integrand in (\ref{Grassmann integral}) is
\begin{displaymath}
\frac{1}{\sqrt{(2E+2Rck)-(c^2+2Rck)\tanh^{2}kx}},
\end{displaymath}
and -- viewed as a real function -- has an indefinite integral, the
form of which depends on the constants involved:
\begin{equation}
\int \frac{1}{\sqrt{a-b \tanh^{2}kx}}dx = \frac{1}{k}\left\{
\begin{array}{cl}
\frac{1}{\sqrt{b-a}} \arcsin \left(\sqrt{\frac{b}{a}-1}\, \sinh x\right)
&; \frac{b}{a} > 1\\
\frac{1}{\sqrt{a-b}} \mbox{arcsinh} \left(\sqrt{1-\frac{b}{a}}\, \sinh x
\right)
&; \frac{b}{a} < 1\\
\frac{1}{\sqrt{a}} \sinh x&; \frac{b}{a}=1
\end{array}
\right.
\label{integrals}
\end{equation}
Because the Grassmann-valued function is completely determined by the
corresponding real function we have to look at the body of
\begin{displaymath}
\frac{b}{a}=\frac{c^2+2ckR}{2E+2ckR}
\end{displaymath} 
to decide which integral to use. Since $R$ as product of two
odd Grassmann constants has no body, the crucial term is
$\frac{c_{b}^2}{2E_{b}}$. 

All three integrals (\ref{integrals}) can be inverted to give $x$ as a
function of $t$:
\begin{equation}
x(t)=\frac{1}{k}\left\{
\begin{array}{ll}
\mbox{arcsinh}\left(\sqrt{\frac{2E+2ckR}{c^2-2E}}
\sin(\,\omega_{I}(t\!-\!\tn)\!+\!\kappa_{I})\right)&;c_{b}^2>2E_{b}\\
\mbox{arcsinh}\left(\sqrt{\frac{2E+2ckR}{2E-c^2}}
\sinh(\,\omega_{II}(t\!-\!\tn)\!+\!\kappa_{II})\right)&;c_{b}^2<2E_{b}\\
\mbox{arcsinh}\left(\,\omega_{III}(t\!-\!\tn)\!+\!\kappa_{III}\right)&
;c_{b}^2=2E_{b}
\end{array}
\right.
\label{xnhyperbolgrassmann}
\end{equation}
where
\begin{displaymath}
\begin{array}{lclclcl}
\omega_{I}&=&k\sqrt{c^2\!-\!2E}&,&
\kappa_{I}&=&
\arcsin\left(\sqrt{\frac{c^2-2E}{2E+2ckR}}\sinh(kx(\tn))\right)\\
\omega_{II}&=&k\sqrt{2E\!-\!c^2}&,&
\kappa_{II}&=&
\mbox{arsinh}\left(\sqrt{\frac{2E-c^2}{2E+2ckR}}\sinh(kx(\tn))\right)\\
\omega_{III}&=&k\sqrt{2E\!+\!2ckR}&,&
\kappa_{III}&=&\sinh kx(\tn).
\end{array}
\end{displaymath}

\bigskip
From equations (\ref{fermion1}) and (\ref{fermion2}) we can calculate
the fermion solutions:
\begin{equation}
\begin{array}{lcll}
\left.
\begin{array}{l}
\psi_{+} \\
\psi_{-} 
\end{array}
\right\}&=&\frac{1}{2E}\frac{\sqrt{2E\!+\!2ckR}}{\sqrt{1\!+\!\frac{c^2+2ckR}
{c^2-2E}\tan^2 y_{I}}}&
\left\{
\begin{array}{l}
\pm Q-\tilde{Q}\sqrt{\frac{c^2}{c^2-2E}}\tan y_{I}\\
\pm \tilde{Q}+Q\sqrt{\frac{c^2}{c^2-2E}}\tan y_{I}
\end{array}
\right.
;c_{b}^2>2E_{b},
\\
\left.
\begin{array}{l}
\psi_{+} \\
\psi_{-} 
\end{array}
\right\}\!&=&\!\frac{1}{2E}\frac{\sqrt{2E\!+\!2ckR}}
{\sqrt{1\!+\!\frac{c^2+2ckR}{2E-c^2}\tanh^2 y_{II}}}&
\left\{
\begin{array}{l}
\!\pm Q-\tilde{Q}\sqrt{\frac{c^2}{2E-c^2}}\tanh y_{II}\\
\!\pm \tilde{Q}+Q\sqrt{\frac{c^2}{2E-c^2}}\tanh y_{II}
\end{array}
\right.
;c_{b}^2<2E_{b},
\\
\left.
\begin{array}{l}
\psi_{+} \\
\psi_{-} 
\end{array}
\right\}\!&=&\!\frac{1}{2E}\frac{1}{\sqrt{1\!+\! y_{III}}}&
\left\{
\begin{array}{l}
\!\pm Q\sqrt{2E\!+\!2ckR}-\tilde{Q}c y_{III}\\
\!\pm \tilde{Q}\sqrt{2E\!+\!2ckR}+Qc y_{III}
\end{array}
\right.
;c_{b}^2=2E_{b},
\end{array}
\end{equation}
with $y_{I}= \pm\, \omega_{I}\,(t\!-\!\tn)\!+\!\kappa_{I}$, etc.
Again, we have found the complete solution for the system without
using any information about the Grassmann algebra $\mathcal{B}$
itself. We will return to this example later, after we have investigated a
special class of Grassmann algebras.

\section{Finitely generated Grassmann algebras}
\label{finite algebra}
We will now make a choice for the underlying Grassmann algebra
$\mathcal{B}$ of the system, namely that it is generated by a finite
number of elements \mbox{$\xi_{i}, i=1,\!\dots\!,n$} with the property
\begin{displaymath}
\xi_{i}\xi_{j}=-\xi_{j}\xi_{i}.
\end{displaymath}
Every element $z$ of $\mathcal{B}$ can be written in the
form
\begin{displaymath}
z=\sum_{k=0}^{n}z_{i_{1}\dots i_{k}} \xi_{i_{1}}\dots \xi_{i_{k}}
\end{displaymath}
with complex coefficients $z_{i_{1}\dots i_{k}}$. $z_{0}$ is the body
of $z$. The requirement for $z$ to be real (in the sense explained
earlier) fixes each coefficient to be either purely real or purely
imaginary, depending on the number of generators involved. For
practical reasons we will usually restrict ourselves to the case of
four generators, although the general aspects of our discussion should
be true for an arbitrary number $n$.

In the case of four generators it is useful to define the following
real monomials: 
\begin{displaymath}
\xi_{ij}=i \xi_{i}\xi_{j},\: \xi_{ijk}=i \xi_{i}\xi_{j}\xi_{k},\::
\xi_{1234}=- \xi_{1}\xi_{2}\xi_{3}\xi_{4}.
\end{displaymath}
Note that due to antisymmetry there are only six linearly independent 
monomials involving two generators, four monomials involving three
generators, and just one monomial of highest order, i.e. involving all
four generators. 

We can decompose the dynamical quantities \mbox{$x$, $\psi_{+}$}
\mbox{and $\psi_{-}$} as follows: 
\begin{displaymath}
\begin{array}{lcl}
x(t)&=&\xn(t)+\xij(t)\xiij+x_{1234}(t)\xi_{1234}\\
\psi_{+}(t)&=&\li(t)\xi_{i}+\lijk(t)\xiijk\\
\psi_{-}(t)&=&\ri(t)\xi_{i}+\rijk(t)\xiijk,
\end{array}
\end{displaymath}
where it is implied from here on that the summation over indices
is with $i<j<k$. In total there are 24 independent real functions which 
we call of zeroth, first, second, third or fourth order according to 
the number of generators involved in the corresponding monomials.

Every analytic function involving $x$, $\psi_{+}$ or $\psi_{-}$ can be
decomposed as well using its Taylor expansion. Applied to the
potential function $U$ this yields:
\begin{displaymath}
U(x)=U(\xn)+U'(\xn)\xij\xiij+\left(U'(\xn)x_{1234}+\frac{1}{2}U''(\xn)
x_{[12}x_{34]}\right)\xi_{1234},
\end{displaymath}
where the brackets denote antisymmetrization. For example
\begin{displaymath}
x_{[12}x_{34]}=
x_{12}x_{34}-x_{13}x_{24}+x_{14}x_{23}+x_{23}x_{14}-x_{24}x_{13}+x_{34}x_{12}.
\end{displaymath}
Note that $U(\xn)$ and its derivatives are ordinary real functions.
Inserting these results into (\ref{Lagrangian}) yields eight component
Lagrangians of order zero, two and four, respectively:
\begin{samepage}
\begin{eqnarray}
L_{0} & = & \frac{1}{2}\dot{\xn}^2-\frac{1}{2}U^2\label{L0}\\
L_{ij} & = & \dot{x}_{0}\dot{x}_{ij}-UU'\xij+\li\dot{\lambda}_{i}+
\ri\dot{\rho}_{i}+U'(\li\rho_{j}-\ri\lambda_{j})\label{Lij}\\
L_{1234} & = & \frac{1}{2}\dot{x}_{[12}\dot{x}_{34]}+\dot{\xn}\dot{x}_{1234}-
\frac{1}{2}\left(UU'\right)'x_{[12}x_{34]}-UU'x_{1234}\label{L1234}\\
&+&\lambda_{[123}\dot{\lambda}_{4]}\!+\!\rho_{[123}\dot{\rho}_{4]}+\!U'
\left(\lambda_{[123}\rho_{4]}\!-\!\rho_{[123}\lambda_{4]}\right)+U''x_{[12}
\left(\lambda_{3}\rho_{4}\!-\!\lambda_{4}\rho_{3}\right)_{]}
\nonumber
\end{eqnarray}
\end{samepage}
where the argument of $U$, $U'$ and $U''$ is always $\xn(t)$, here and
below.

From the highest order Lagrangian $L_{1234}$, which is a functional of 24
generally different component functions, we get the following
set of Euler-Lagrange equations:
\begin{eqnarray}
\ddot{x}_{0} & = & - UU'\label{x0}\\
\ddot{x}_{ij} & = & - (UU')'\xij+U''(\li\rho_{j}-\ri\lambda_{j})\label{xij}\\
\begin{array}{l}
\ddot{x}_{1234}\\
\mbox{}
\end{array}
&
\begin{array}{c}
=\\
\mbox{}
\end{array}
&
\begin{array}{l}
 -\frac{1}{2}(UU')'' x_{[12}x_{34]} - (UU')'x_{1234}\\
+U''\left(\lambda_{[123}\rho_{4]}-\rho_{[123}\lambda_{4]}\right)
+U'''x_{[12}\left(\lambda_{3}\rho_{4}-\lambda_{4}\rho_{3}\right)_{]}
\end{array}\label{x1234}
\\
\left.
\begin{array}{l}
\dot{\lambda}_{i}\\
\dot{\rho}_{i}
\end{array}
\right\}&=&
\left\{
\begin{array}{l}
- U' \ri\\
\;\;\,U' \li
\end{array}
\right.\label{liri}\\
\left.
\begin{array}{l}
\dot{\lambda}_{ijk}\\ 
\dot{\rho}_{ijk}
\end{array}
\right\}
&=&
\left\{
\begin{array}{l}
- U' \rijk - U''x_{[ij}\rho_{k]}\\
\;\;\, U' \lijk + U''x_{[ij}\lambda_{k]}
\end{array}
\right.\label{lijkrijk}
\end{eqnarray}
The same equations can also be obtained by splitting the original
equations of motion (\ref{x-equation})--(\ref{fermion-equation 2}) into
their components.

It is remarkable that all equations can be derived from just the one
Lagrangian. The other seven component Lagrangians are completely redundant
and yield no new equations: \mbox{From (\ref{L0})} we can only obtain
equation (\ref{x0}) whereas from the Lagrangians of type (\ref{Lij})
we can derive equations (\ref{x0}), (\ref{xij}) and (\ref{liri}).

\bigskip
Now we can also see where the ambiguity mentioned earlier comes from
when there is an odd number $n$ of generators: Because the Lagrangian
is an even functional, none of its components can contain any function
of highest order $n$, so these functions cannot be governed  by any
equations of motion. E.g. for three generators the functions
$\lambda_{123}$ and $\rho_{123}$ will stay completely undetermined. 
Formally, we can derive evolution equations similar to (\ref{lijkrijk})
by decomposing (\ref{fermion-equation 1}) and (\ref{fermion-equation
2}) into components but they are meaningless since an arbitrary
function can be added to both sides. Thus we can regard these highest
order functions as non-physical and treat them as trivially separated gauge
degrees of freedom.
However, if there is an even number of generators this problem cannot
occur, since there will always be one Lagrangian of highest order
containing component functions of all orders which are
thus determined by equations of motion.

\section{Layer-by-layer solutions}
\label{solarbpot}
We now proceed to derive the solutions to the equations of
motion (\ref{x0})--(\ref{lijkrijk}) for an arbitrary, sufficiently
differentiable potential function $U$ adopting a layer-by-layer
strategy. That means that we will start with the lowest order or
'body' equation (\ref{x0}) and then use its solution to work our way up
to the higher order and more complex equations, including also the
fermionic ones. We do this here up to fourth order, although there is
no obstruction in principle to continue the strategy for an algebra
with more than four generators. In fact, in every layer the equations and their
solutions are the same irrespective of how many generators there are;
there will only be a different number of them.

\bigskip
Before we start with the bosonic equations, it is useful to look at the 
decomposition of the Hamiltonian
$H=H_{0}+H_{ij}\xiij+H_{1234}\xi_{1234}$ and of the Noether charge 
$R=R_{ij}\xiij+R_{1234}\xi_{1234}$ (note that $R$ does not have a
body):  
\begin{eqnarray}
H_{0}&=&\frac{1}{2}\dot{x}_{0}^2+\frac{1}{2}U^{2}(\xn)\label{H0}\\
H_{ij}&=&\dot{x}_{0}\dot{x}_{ij}+UU'\xij-U'\Rij\label{Hij}\\
H_{1234}&=&\frac{1}{2}\dot{x}_{[12}\dot{x}_{34]}+\dot{x}_{0}\dot{x}_{1234}+
\frac{1}{2}\left(UU'\right)'x_{[12}x_{34]}+UU'x_{1234}\\
&&-U'\Rtwo-
U''x_{[12}R_{34]}\nonumber\\
\Rij&=&\li\rho_{j}-\ri\lambda_{j}\label{Rij}\\
\Rtwo&=&\lambda_{[123}\rho_{4]}-\rho_{[123}\lambda_{4]}.\label{R1234}
\end{eqnarray}
Since these are components of Grassmann conserved quantities,
they are conserved too, as can be checked easily using
the equations of motion. All components can be derived as Noether
charges from the highest order Lagrangian $L_{1234}$ and we will do
this in section \ref{symmetries} but for now we only need that they
are constant in time. The respective values of the Hamiltonian
functions $H_{0}$, $H_{ij}$ and $H_{1234}$ are denoted in the
following by $E_{0}$, $E_{ij}$ and $E_{1234}$.

\bigskip
We notice first that (\ref{x0}) is just
the standard Newtonian equation of motion for a particle moving in a
potential $\frac{1}{2}U^2$. This is the bottom layer of the system.

The solution to (\ref{x0}) using the constancy of the Hamiltonian
$H_{0}$ is well-known: 
\begin{equation}
t-\tn = \pm
\int_{\xn(\tn)}^{\xn(t)}\frac{1}{\sqrt{2E_{0}-U^2(\xn')}}d\xn'.\label{solutxn}
\end{equation}
The sign has to be chosen carefully to comply with the direction of
motion of the particle; if the particle motion changes direction, the
integral will be multi-valued and has to be glued together from pieces
with unique sign to ensure that the overall result for  $t(\xn)$ is
monotonically growing. The implicit function theorem then allows us to locally
invert $t(\xn)$ and obtain the required $\xn(t)$. Since (\ref{x0}) is
a second order equation there are two constants of integration,
$\xn(\tn)$ (or equivalently $\tn$) and the energy $E_{0}$.

\bigskip
Whereas (\ref{x0}) is a non-linear equation, (\ref{xij}) is an
inhomogeneous \textit{linear} equation (which can be simplified using
(\ref{Rij})). It is now convenient to use the Hamiltonian (\ref{Hij}): 
Solving the equation $H_{ij}=E_{ij}$ for $\dot{x}_{ij}$ we can reduce 
the problem to a first order differential equation. This can be solved 
by standard methods to yield:
\begin{displaymath}
\xij=c_{ij}\dot{x}_{0}+E_{ij}\dot{x}_{0}\int_{\tn}^t\frac{1}
{\dot{x}_{0}^2}dt'+\Rij\dot{x}_{0}\int_{\tn}^t\frac{U'}{\dot{x}_{0}^2}dt'.
\end{displaymath}
where $c_{ij}$ is an arbitrary integration constant related to the
initial value of $\xij$. The second integration constant
of (\ref{xij}) is the energy variable $E_{ij}$. This result was also
found in \cite{Junker} with $R_{ij}$ set to $-1$.  
However, using equation (\ref{x0}) and $H_{0}=E_{0}$, the last term on the
right hand side can be rewritten:
\begin{equation}
\Rij\dot{x}_{0}\int_{\tn}^t\frac{U'}{\dot{x}_{0}^2}dt'=
\Rij\dot{x}_{0}\int_{\tn}^t\frac{U'(\dot{x}_{0}^{2}+U^{2})}
{2E_{0}\dot{x}_{0}^2}dt'=
\frac{\Rij}{2E_{0}}\dot{x}_{0}\int_{\tn}^t\frac{\dot{U}\dot{x}_{0}-
U\ddot{x}_{0}}{\dot{x}_{0}^2}dt'=
\frac{R_{ij}}{2E_{0}}U.
\label{Rijrewritten}
\end{equation}
so that we end up with the simpler expression
\begin{equation}\label{solutxij}
\xij=c_{ij}\dot{x}_{0}+E_{ij}\dot{x}_{0}\int_{\tn}^t\frac{1}
{\dot{x}_{0}^2}dt'+\frac{\Rij}{2E_{0}}U.
\end{equation}
All three functions which occur in (\ref{solutxij}) can be
calculated directly from $\xn(t)$.

\bigskip
The same procedure can be applied to solve (\ref{x1234}) which
simplifies when we use (\ref{Rij}) \mbox{and (\ref{R1234})}. We rearrange the
equation $H_{1234}=E_{1234}$ to isolate $\dot{x}_{1234}$ and then solve
this first order equation.
The result can be written as:
\begin{eqnarray}
x_{1234}&=&c_{1234}\dot{x}_{0}+E_{1234}\dot{x}_{0}\int_{\tn}^t\frac{1}
{\dot{x}_{0}^2}dt'+\frac{R_{1234}}{2E_{0}}U\label{solutx1234}\\
&&+\frac{1}{2}\dot{x}_{0}\int_{\tn}^t\frac{x_{[12}
\ddot{x}_{34]}-\dot{x}_{[12}\dot{x}_{34]}}{\dot{x}_{0}^2}dt'
+\frac{1}{2}\dot{x}_{0}\int_{\tn}^t\frac{U''x_{[12}R_{34]}}{\dot{x}_{0}^2}dt'
\nonumber.
\end{eqnarray}
Again, there are two additional integration constants, $c_{1234}$
and $E_{1234}$. Note that the solution depends only on functions that
we already know from the lower layers, namely $\xn$ and $\xij$. In
fact, the solution can be expressed as a function of $\xn$ and its
derivatives only by inserting (\ref{solutxij}) into
(\ref{solutx1234}):
\begin{eqnarray}
x_{1234}&=&c_{1234}\dot{x}_{0}+E_{1234}\dot{x}_{0}\int_{\tn}^t\frac{1}
{\dot{x}_{0}^2}dt'+\frac{R_{1234}}{2E_{0}}U\label{4var}\\
&&+\frac{1}{2}c_{[12}c_{34]}\ddot{x}_{0}+c_{[12}E_{34]}\left(
\ddot{x}_{0}\int_{\tn}^t\frac{1}
{\dot{x}_{0}^2}dt'+\frac{1}{\dot{x}_{0}}\right)\nonumber\\
&&+\frac{1}{2}E_{[12}E_{34]}\left[\ddot{x}_{0}\left(\int_{\tn}^t\frac{1}
{\dot{x}_{0}^2}dt'\right)^2+\frac{2}{\dot{x}_{0}}\int_{\tn}^t\frac{1}
{\dot{x}_{0}^2}dt'-3\dot{x}_{0}\int_{\tn}^t\frac{1}{\dot{x}_{0}^4}dt'
\right]\nonumber\\
&&+\frac{R_{[12}c_{34]}}{2E_{0}}U'\dot{x}_{0}+\frac{R_{[12}E_{34]}}{2E_{0}}
\left(U'\dot{x}_{0}\int_{\tn}^t\frac{1}{\dot{x}_{0}^2}dt'
-\frac{U}{E_{0}}\right)
\nonumber.
\end{eqnarray}
This complicated-looking expansion is particularly useful when
we come to the interpretation of the solution in section
\ref{interpret}.

\bigskip
There remain the fermion equations (\ref{liri}) and (\ref{lijkrijk})
to be solved. Both are systems of two homogeneous first order equations and
therefore have two independent solutions. Just as we used the decomposition
of $H$ to solve the bosonic equations, so it is now appropriate to
decompose the conserved supercharges $Q=Q_{i}\xi_{i}+Q_{ijk}\xiijk$ and
$\tilde{Q}=\tilde{Q}_{j}\xi_{i}+\tilde{Q}_{ijk}\xiijk$ to solve the fermionic
equations. We find:
\begin{equation}
\label{qiqti}
\begin{array}{lcl}
Q_{i}&=&\li\dot{x}_{0}+\ri U,\\
\tilde{Q}_{i}&=&\ri\dot{x}_{0}-\li U,\\
Q_{ijk}&=&\lijk\dot{x}_{0}+\rijk U+\lambda_{[i}\dot{x}_{jk]}
+\rho_{[i}U'x_{jk]},\\
\tilde{Q}_{ijk}&=&\rijk\dot{x}_{0}-\lijk U+\rho_{[i}\dot{x}_{jk]}
-\lambda_{[i}U'x_{jk]}. 
\end{array}
\end{equation}
Using suitable linear combinations of $\dot{x}_{0}$ and $U$ with
coefficients $Q_{i}$ and $\tilde{Q}_{i}$ and applying $H_{0}=E_{0}$ we
find that
\begin{equation}
\begin{array}{lcl}
\li=l_{i}\dot{x}_{0} -r_{i}U,\:\:\:\:
\ri=r_{i}\dot{x}_{0} +l_{i}U,
\end{array}
\label{solutliri}
\end{equation}
where
$l_{i}=\frac{Q_{i}}{2E_{0}}$ and $r_{i}=\frac{\tilde{Q}_{i}}{2E_{0}}$;
hence the two integration constants of the solution are basically
given by the first order supercharges. A similar result was obtained
in \cite{Junker}, although the restrictions placed there on the shape of the
potential $U$ are unnecessary.  

Note, however, that this solution does not work if the
particle energy $E_{0}$ equals zero. This has to do with the fact that
both $\dot{x}_{0}$ and $U$ must vanish in this case. We will return to
this in section \ref{zeroenergy}.

\bigskip
The same procedure can be used to find the solution to equation
(\ref{lijkrijk}) and we get the result:
\begin{eqnarray} 
\lijk&=&l_{ijk}\dot{x}_{0}-r_{ijk}U-\frac{E_{[ij}\lambda_{k]}}{2E_{0}}
+\frac{U\dot{x}_{[ij}\rho_{k]}-U'\dot{x}_{0}x_{[ij}\rho_{k]}}{2E_{0}}
\label{solutlijkrijk1}\\
\rijk&=&r_{ijk}\dot{x}_{0}+l_{ijk}U-\frac{E_{[ij}\rho_{k]}}{2E_{0}}
-\frac{U\dot{x}_{[ij}\lambda_{k]}-U'\dot{x}_{0}x_{[ij}\lambda_{k]}}{2E_{0}},
\label{solutlijkrijk2}
\end{eqnarray}
where we have used the equations $H_{0}=E_{0}$ and $H_{ij}=E_{ij}$ and
the identities $R_{[ij}\lambda_{k]}=$\mbox{$R_{[ij}\rho_{k]}=0$}. The
new integration constants $l_{ijk}$ and $r_{ijk}$ are related to the
third order supercharges $Q_{ijk}$ and $\tilde{Q}_{ijk}$ in the same
way as $l_{i}$ and $r_{i}$ are to $Q_{i}$ and $\tilde{Q}_{i}$. As was
the case for the second and fourth order bosonic solutions we find
that $\lijk$ and $\rijk$ depend only on functions we know already from
the lower order layers, namely, $\xn$, $\li$, $\ri$ and $\xij$, and
again we can simplify the solution even further, inserting
(\ref{solutxij}) and (\ref{solutliri}) into equations
(\ref{solutlijkrijk1}) and (\ref{solutlijkrijk2}):
\begin{samepage}
\begin{eqnarray}
\lijk&=&\l_{ijk}\dot{x}_{0}+c_{[ij}l_{k]}\ddot{x}_{0}+E_{[ij}l_{k]}
\left(\ddot{x}_{0}\int_{\tn}^t\frac{1}{\dot{x}_{0}^2}dt'+\frac{1}{\dot{x}_{0}}-
\frac{\dot{x}_{0}}{E_{0}}\right)\label{3rdvar1}\\
&&-r_{ijk}U-c_{[ij}r_{k]}U'\dot{x}_{0}-E_{[ij}r_{k]}
\left(U'\dot{x}_{0}\int_{\tn}^t\frac{1}{\dot{x}_{0}^2}dt'-
\frac{U}{E_{0}}\right)\nonumber\\
\rijk&=&r_{ijk}\dot{x}_{0}+c_{[ij}r_{k]}\ddot{x}_{0}+E_{[ij}r_{k]}
\left(\ddot{x}_{0}\int_{\tn}^t\frac{1}{\dot{x}_{0}^2}dt'+\frac{1}{\dot{x}_{0}}-
\frac{\dot{x}_{0}}{E_{0}}\right)\label{3rdvar2}\\
&&+l_{ijk}U+c_{[ij}l_{k]}U'\dot{x}_{0}+E_{[ij}l_{k]}
\left(U'\dot{x}_{0}\int_{\tn}^t\frac{1}{\dot{x}_{0}^2}dt'-
\frac{U}{E_{0}}\right)\nonumber.
\end{eqnarray}
\end{samepage}
We have now derived all solutions to equations (\ref{x0})--(\ref{lijkrijk}),
i.e. we have explicitly solved the equations of motion up to
fourth order for an arbitrary potential function $U$. They are all
functions of only four quantities and their time derivatives:
\begin{displaymath}
\dot{x}_{0},\:U,\:\int_{\tn}^{t}\frac{1}{\dot{x}_{0}^{2}}dt'\:\:\:
\mbox{and}\:\:\int_{\tn}^{t}\frac{1}{\dot{x}_{0}^{4}}dt'.
\end{displaymath}
To illustrate the method we choose the harmonic oscillator as our first
example. This has the advantage that we can compare the solution with our 
earlier result (\ref{hx-solution})--(\ref{hpsi2-solution}).

\subsection{The harmonic potential $U(x)=kx$}
\label{harmpot}
The solution for $\xn$ could be found using (\ref{solutxn}) but here
it is easy to solve equation (\ref{x0}) directly, with the
familiar result:
\begin{equation}
\xn(t)=\xn(\tn)\cos k(t\!-\!\tn)+\frac{v_{0}}{k}\sin
k(t\!-\!\tn);\:v_{0}=\pm\sqrt{2E_{0}\!-\!k^2 x_{0}^{2}(\tn)}.
\end{equation}
The functions $\dot{x}_{0}$ and $U$ can be calculated easily
and we find:
\begin{eqnarray}
\li(t)&=&L_{i}\cos k(t\!-\!\tn)-R_{i}\sin k(t\!-\!\tn),\label{fermionharm1}\\
\ri(t)&=&L_{i}\sin k(t\!-\!\tn)+R_{i}\cos k(t\!-\!\tn),\label{fermionharm2}
\end{eqnarray}
where $L_{i}$ and $R_{i}$ are the constants
\begin{eqnarray}
L_{i}=\frac{Q_{i}}{2E_{0}}\!v_{0}\!-\!\frac{\tilde{Q_{i}}}{2E_{0}}\!k\xn(\tn),
\label{Li}\\
R_{i}=\frac{Q_{i}}{2E_{0}}\!kx_{0}(\tn)\!
+\!\frac{\tilde{Q_{i}}}{2E_{0}}\!v_{0}.
\label{Ri}
\end{eqnarray}
Next we have to calculate $\xij$ and thus we need
\begin{equation}
\label{harmxij}
\int_{\tn}^{t} \frac{1}{\dot{x}_{0}^{2}}dt' = \int_{\xn(\tn)}^{\xn(t)}
\frac{1}{2E_{0}-k^2\xn'^2}d\xn'=\frac{1}{2E_{0}}
\left(\frac{\xn(t)}{\dot{x}_{0}(t)}-\frac{\xn(\tn)}{v_{0}}\right).
\end{equation}
On using (\ref{solutxij}) we find that
\begin{eqnarray}
\xij(t)&=&\left(\frac{\Rij}{2E_{0}}k\xn(\tn)\!+\!c_{ij}v_{0}\right)
\cos k(t\!-\!\tn)+ \left(\frac{\Rij}{2E_{0}}v_{0}\!-\!c_{ij}k\xn(\tn)\!+\!
\frac{E_{ij}}{kv_{0}}\right) \sin k(t\!-\!\tn).\label{harmxijsolut}
\end{eqnarray}
For $\lijk$ and $\rijk$ we need not calculate any new terms except
$\ddot{x}_{0}$. After collecting various constants into $L_{ijk}$ and
$R_{ijk}$ we end up with
\begin{eqnarray*}
\lijk&=&L_{ijk}\cos k(t\!-\!\tn) - R_{ijk}\sin k(t\!-\!\tn)\\
\rijk&=&L_{ijk}\sin k(t\!-\!\tn) + R_{ijk}\cos k(t\!-\!\tn).
\end{eqnarray*} 
Finally we have to calculate $x_{1234}$. Therefore we need the
integral
\begin{displaymath}
\int_{\tn}^{t}\frac{1}{\dot{x}_{0}^{4}}dt'=\left[\frac{1}{6E_{0}}
\frac{\xn}{\dot{x}_{0}^3}+\frac{1}{6E_{0}^2}\frac{\xn}{\dot{x}_{0}}
\right]_{\xn(\tn)}^{\xn(t)}.
\end{displaymath}
There seems to be a singularity when $\dot{x}_{0}$ approaches zero but
fortunately this term has to be multiplied by $\dot{x}_{0}$ and
combined with two other terms which share the same coefficient 
$E_{[12}E_{34]}$. One finds that all terms involving
$\dot{x}_{0}$ in the denominator cancel each other, so that the overall
result is just a linear combination of $\xn$ and $\dot{x}_{0}$,
i.e. of $\cos$- and $\sin$-terms. This holds also for all other
contributions to $x_{1234}$, so that we can write:
\begin{displaymath}
x_{1234}=C_{1}\cos k(t\!-\!\tn) + C_{2}\sin k(t\!-\!\tn),
\end{displaymath}
where $C_{1}$ and $C_{2}$ are rather complicated functions of
$E_{1234}$, $c_{1234}$ and all lower order integration constants.

The most interesting observation is that all bosonic \textit{and}
fermionic functions are linear combinations of $\cos$- and
$\sin$-terms only. As we will see later, this similarity between the 
solutions of different orders is the exception rather than the rule 
and a special feature of the harmonic oscillator. We also find that 
the fermionic functions oscillate with the same period as the 
bosonic functions and differ only in amplitude and phase. Once the
motion of the lowest order bosonic function $\xn$ is fixed, the
principal motion for $\li$, $\ri$ is almost completely determined. 
This is a general feature of our system, and it can be fully understood 
by investigating the symmetries of our model.

Finally, it is easy to verify that the complete component solution is 
compatible with the general solution found in section \ref{harmonicpotential}.

\section{Symmetries in component form}
\label{symmetries}
We have already dealt with the decomposition of the Hamiltonian $H$, the
constant $R$ and the two supercharges $Q$ and $\tilde{Q}$. Now we want
to explain the origin of the component charges from the underlying
component symmetries. The Lagrangian $L_{1234}$ (\ref{L1234}) is all
we need to look at, since the non-trivial symmetries of all lower
order Lagrangians form subgroups of the full symmetry group of
$L_{1234}$. One would not guess the extraordinary number of symmetries
of $L_{1234}$ if one did not know their supersymmetric origin.

\bigskip
To start with, (\ref{L1234}) is invariant under time translations, and
this leads to the conservation of the highest order Hamiltonian
$H_{1234}$. This result can be generalized by looking at the full
Grassmann transformation (\ref{timetranslation}). Decomposion into
components gives us back the time translation symmetry mentioned above
when $\eta=\eta_{0}$ is real; the choices $\eta=\eta_{ij}\xi_{ij}$ (no
summation) or $\eta=\eta_{1234}\,\xi_{1234}$ lead to seven extra charges
which correspond to the Hamiltonians $H_{ij}$ and $H_{0}$, respectively.

\bigskip
Apart from the Hamiltonians, sixteen Noether charges $Q_{i},\,
\tilde{Q}_{i},\, Q_{ijk}$ and $\tilde{Q}_{ijk}$ derive from the 
supercharges $Q$ and $\tilde{Q}$ and belong to a set of symmetry
transformations which can be obtained from (\ref{SUSY transformation})
by the choices $\epsilon=\epsilon_{i}\xi_{i}$ and
$\epsilon=\epsilon_{ijk}\xi_{ijk}$ (no summation both times),
respectively. As an example, $Q_{i}$ and $\tilde{Q}_{i}$ are the
charges that belong to: 
\begin{displaymath}
\begin{array}{lclclclclcl}
\delta_{i}x_{1234}&=&\epsilon_{i} \lambda_{i}&,&
\delta_{i}\lambda_{jkl}&=&-\epsilon_{i}\dot{x}_{0}&,&
\delta_{i}\rho_{jkl}&=&-\epsilon_{i} U,\\
\tilde{\delta}_{i}x_{1234}&=&\epsilon_{i}\rho_{i}&,&
\tilde{\delta}_{i}\lambda_{jkl}&=&\epsilon_{i} U&,&
\tilde{\delta}_{i}\rho_{jkl}&=&-\epsilon_{i} \dot{x}_{0},
\end{array}
\end{displaymath}
where $\{ijkl\}$ is a cyclic permutation of $\{1234\}$ and no
summation over $i$ is implied. The term supersymmetry transformations
should be avoided here, though, since the dynamical quantities in
component form are just real functions -- therefore they can be termed
bosonic or fermionic only by convention. Notwithstanding this fact the
component transformations and the corresponding charges
\textit{reflect} the underlying supersymmetry.

\bigskip
Further component Noether charges derived from the Lagrangian (\ref{L1234})
are $R_{ij}$ and $R_{1234}$. The transformations can be read off from
(\ref{fermion rotation}) choosing $\eta=\eta_{ij}\,\xi_{ij}$ (no
summation) and \mbox{$\eta=\eta_{0}$ (real)}, respectively. In the
latter case we get, for example  
\begin{equation}
\label{fermion rotation component}
\begin{array}{lclclcl}
\delta_{1234}\lambda_{i}&=&\eta_{0}\, \rho_{i}&,&
\delta_{1234}\rho_{i}&=&-\eta_{0}\, \lambda_{i},\\
\delta_{1234}\lambda_{ijk}&=&\eta_{0}\, \rho_{ijk}&,&
\delta_{1234}\rho_{ijk}&=&-\eta_{0}\, \lambda_{ijk}.
\end{array}
\end{equation}

\bigskip
So far all transformations have just resembled those found in section
\ref{grassmannsymmetries}. There are however further sets of symmetry
transformations which only occur for the component Lagrangian $L_{1234}$.
The easiest of these is given by
\begin{equation}
\label{lintrafo1}
\delta_{i}\lambda_{jkl}=-\eta_{i}\lambda_{i},\;\delta_{i}\rho_{jkl}=
-\eta_{i}\rho_{i},
\end{equation}
where again $\{ijkl\}$ is a cyclic permutation of $\{1234\}$. This
leads to four charges 
\begin{equation} 
\label{lincharge1}
S_{i}=\frac{1}{2}(\lambda_{i}^2+\rho_{i}^2).
\end{equation}
The next easiest is
\begin{equation}
\label{lintrafo2}
\delta_{ij}\lambda_{ikl}=\eta_{ij}\lambda_{i},\;
\delta_{ij}\lambda_{jkl}=-\eta_{ij}\lambda_{j},\;
\delta_{ij}\rho_{ikl}=\eta_{ij}\rho_{i},\;
\delta_{ij}\rho_{jkl}=-\eta_{ij}\rho_{j}
\end{equation}
and leads to six independent charges
\begin{equation}
\label{lincharge2}
S_{ij}=\lambda_{i}\lambda_{j}+\rho_{i}\rho_{j}.
\end{equation}
We shall exploit these charges later but for now we are more
interested in where they come from. It turns out that 
these and two other groups of transformations not mentioned yet reflect
invariance of the original Lagrangian under change of basis in the
Grassmann algebra. A Grassmann algebra with $n$ generators can be
viewed as a $2^{n}$-dimensional vector space in the obvious way, but the
choice of the generators $\xi_{i}$ and their products as basis vectors
is somewhat arbitrary. A change of basis, however, has to be compatible 
with the multiplicative structure of the algebra. This means firstly 
that we only have to consider transformations of the $n$ generators and 
secondly that we can change an (odd) generator only by another odd 
element of the algebra. The final constraint is that the highest 
order monomial, e.g. $\xi_{1234}$ in the case of four generators, 
remains unchanged by the transformation so that the highest order 
Lagrangian, here $L_{1234}$, remains invariant.
 
In the four generator case, this leaves only a relatively small group of
acceptable linear transformations. There are eight independent
odd basis elements, namely $\xi_{i}$ and $\xi_{ijk}$. Correspondingly,
there are the following independent infinitesimal transformations that
fulfil our conditions:
\begin{enumerate}
\item $\xi_{i} \mapsto \xi_{i} + \eta\xi_{j}\; (i\neq j)$\label{active3}
\item $\xi_{i} \mapsto \xi_{i}-\eta\xi_{jkl}\; (\{ijkl\}$ an even 
permutation of $\{1234\}$)\label{active1} 
\item $\xi_{i} \mapsto \xi_{i}+\eta \xi_{i}, \:\xi_{j} \mapsto \xi_{j} -
\eta \xi_{j}\;(i\neq j)$: There are just three independent \textit{scaling}
transformations of this type. \label{active4} 
\item $\xi_{i} \mapsto \xi_{i}+\eta\xi_{ikl},\: \xi_{j} \mapsto
\xi_{j}-\eta\xi_{jkl}\;(\{ijkl\}$ an even permutation of
$\{1234\}$): There are six independent \textit{Grassmann scaling}
transformations of this type.\label{active2}
\end{enumerate}
In order for the Grassmann quantities $x$, $\psi_{+}$ and $\psi_{-}$
to be invariant, corresponding to these transformations of the
\textit{basis vectors} of $\mathcal{B}$ there have to be
transformations of the real \textit{components}. It is these component
transformations that we have found in (\ref{lintrafo1}) and
(\ref{lintrafo2}), which belong to the basis transformations \ref{active1}.
and \ref{active2}., respectively. 

The component transformations belonging to \ref{active3}. and
\ref{active4}. are also symmetries of the Lagrangian,
but since they are more complicated to write down and we will not use
them later we refrain from giving them here. It should be mentioned
though that they generate a full $SL_{4}$ symmetry group.

\section{Bosonic and fermionic motion -- general properties}\label{interpret}
We finally come to the physical interpretation of the results we have
so far obtained concerning the component dynamical variables.
Starting with the lowest order bosonic function $\xn$
we first recall that this always describes the motion of a particle in
one dimension in the potential $\frac{1}{2}U^2(x_{0})$. There are three
possible types of motion in one dimension:
\begin{enumerate}
\item Movement with no turning point: The particle velocity is always
positive (or negative), the motion can be bounded, approaching finite
values of $\xn$ as $t$ goes to $\pm\infty$, or unbounded. Example:
Flat potential function $U(x)=c$.
\item Movement with one turning point: The particle velocity changes
sign once, when $\xn$ reaches a maximal (minimal) value. Again, the
motion can be bounded or unbounded. Example: Reciprocal potential
$U(x)=\frac{c}{x}$.
\item Oscillatory motion: The particle velocity changes sign
infinitely often, thus the motion is always restricted to a finite
interval. Example: Harmonic oscillator $U(x)=kx$.
\end{enumerate}
There is one conserved quantity, namely the lowest order Hamiltonian
$H_{0}$, given in (\ref{H0}). We can interpret this as half the
squared length of the two-dimensional bosonic vector 
\begin{equation}
\left(
\begin{array}{c}
\dot{x}_{0}\\
U
\end{array}
\right).
\label{bosvector}
\end{equation}
The conservation of $H_{0}$ means that the motion of this vector is
restricted to a circle with squared radius $2E_{0}$.

Corresponding to the three types of motion for $\xn$ there are three
types of motion for this vector:

\begin{enumerate}
\item Movement with no turning point: The vector moves on the
right (left) semicircle in the $\dot{x}_{0}$-$U$-plane. The number of
direction changes depends on the  number of potential extrema of
$U$.
\item Movement with one turning point: Motion starts on the right
(left) semicircle of the $\dot{x}_{0}$-$U$-plane, then changes when
the turning point is reached to the other semicircle where it then
mirrors the previous motion.  
\item Oscillatory motion: There are two subcases, depending
on whether the number of sign changes of $U(x)$ between the two
extremal points $x_{min}$ and $x_{max}$ is odd or even.
\begin{enumerate}
\item When there is an odd number the vector will continually wind around the
circle; direction changes depending on the number of extrema of $U$ can
be superimposed. This is the only case where the motion covers the
complete circle. The most important example is the harmonic
oscillator where we can see directly the circular motion of the
bosonic vector. 
\item When the number of sign changes of $U$ is even, the bosonic
vector will oscillate on the circle between two points lying
symmetrically on either side of the $U$-axis. An example of this case
is the potential function $U(x)=\sqrt{(kx)^2+c}$ with $c>0$, which
leads to a harmonic oscillator potential shifted upwards by $c$.
\end{enumerate}
\end{enumerate}

\bigskip
The bosonic vector is important because it largely determines
the behaviour of the $n$ first order fermionic functions. To see this, 
we look at the $n$ two-dimensional vectors 
\begin{equation}
\left(
\begin{array}{l}
\li\\\ri
\end{array}
\right);\;i=1,\dots ,n.
\end{equation}
The charges $S_{i}$ introduced in (\ref{lincharge1}) guarantee that
the lengths of these vectors are conserved, thus their motions are also
restricted to circles. The squared radii are given by $2S_{i}$, providing us 
with a nice geometrical interpretation of these constants. Furthermore, the
charges $S_{ij}$, given by (\ref{lincharge2}), can be
interpreted as \textit{scalar products} between the $i$th and $j$th vectors,
thus effectively specifying the absolute value of the angle between
them. Hence we can see already that all $n$ vectors must corotate.
This can be confirmed further by the charges $R_{ij}$ calculated in 
(\ref{Rij}). They can be seen as two-dimensional \textit{determinants}, 
fixing the area of the parallelogram defined by the above-mentioned two
vectors. It then follows that the sign of the angle between these
is determined, too.

The other conserved quantities that involve only terms of first
or lower order are the supercharges $Q_{i}$ and $\tilde{Q}_{i}$, given
in (\ref{qiqti}). They couple the fermionic vectors with the bosonic
vector: The charges $Q_{i}$ fix the scalar products, the constants
$\tilde{Q}_{i}$ the determinants between $(\li,\ri)$ and $(\dot{x}_{0},U)$, 
so that all angles between these vectors, including their signs, are determined 
and constant in time. This means that once the bosonic motion is calculated 
and the fermionic initial data is given, we can read off the time evolution of
all fermionic functions since their vectors are rigidly corotating
with the bosonic one.

\begin{figure}[ht]
\begin{center}
\epsfig{file=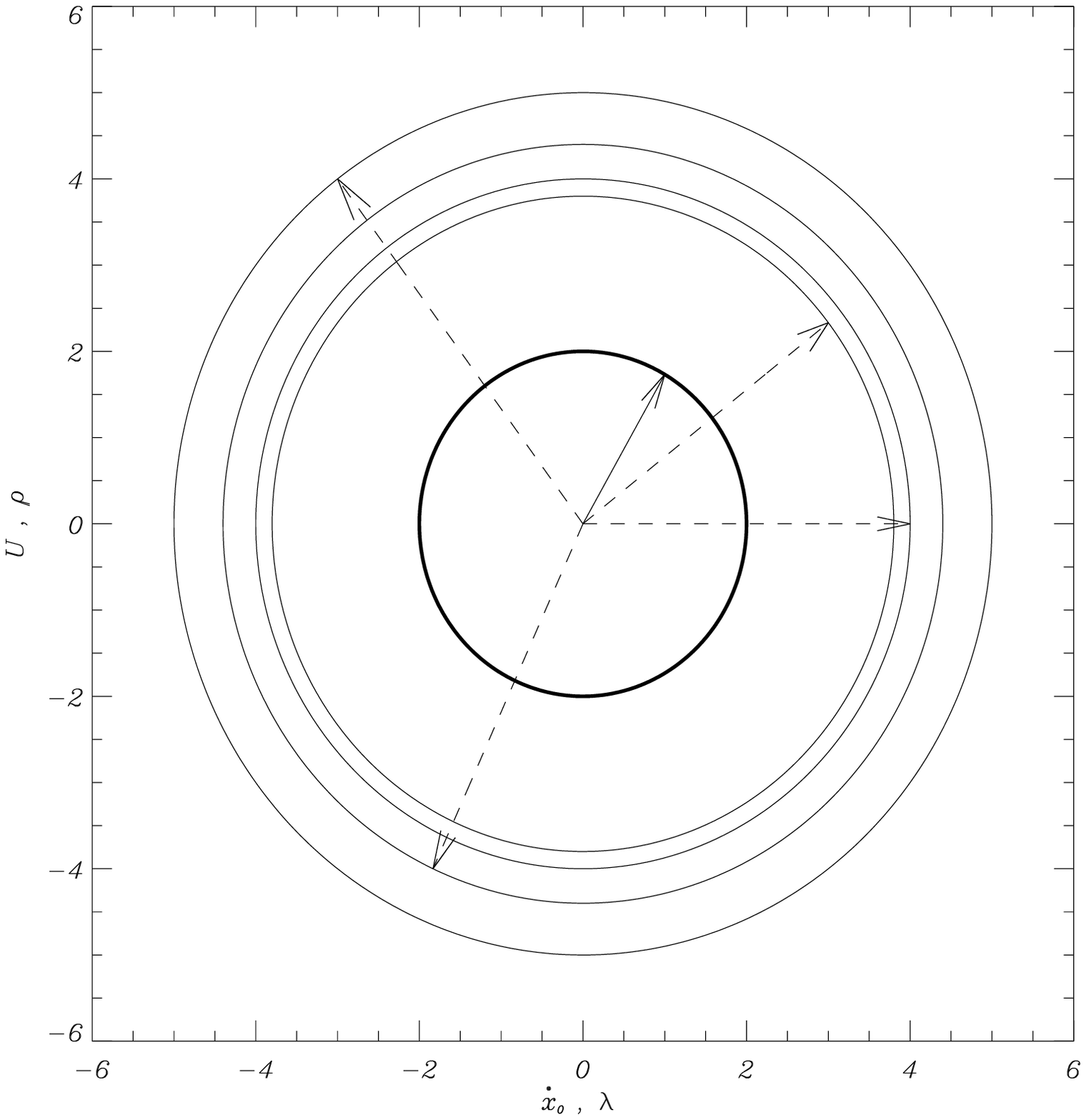,width=7cm}
\caption{{\small \it Bosonic (continuous) and fermionic (dashed) vectors,
rigidly corotating}}
\label{diagram1}
\end{center}
\end{figure}

\bigskip
Having now dealt with the lowest order bosonic and fermionic
functions we have to ask how to interpret the higher order
functions. Unfortunately, the geometric interpretation of these can
not be seen as clearly. Instead we will interpret the higher order
functions as \textit{variations} of those of lower order.

To start with, $\xij$ as given in (\ref{solutxij}) has two terms coming from
the two homogeneous solutions of the $\xij$-equation with coefficients
$c_{ij}$ and $E_{ij}$ and the term with coefficient $\Rij$ which
constitutes a particular solution. We will now show that the first
two terms can be written as variations of the lowest order motion
$\xn$ with respect to the two free parameters $\tn$ and $E_{0}$ of
that motion.

The variation of $\xn(t)$ with $\tn$ is clearly proportional to the
\mbox{velocity $\dot{x}_{0}$:} 
\begin{displaymath}
\frac{\delta \xn}{\delta \tn} =- \dot{x}_{0}
\end{displaymath}
which gives us the term with coefficient $c_{ij}$. Next we vary the
equation $H_{0}=E_{0}$:
\begin{displaymath}
2\dot{x}_{0}\delta\dot{x}_{0}+2UU'\delta x_{0}=2\, \delta E_{0}.
\end{displaymath}
Using equation (\ref{x0}), dividing by $\dot{x}_{0}^2$ and integrating
between $\tn$ and $t$ we obtain 
\begin{displaymath}
\int_{\tn}^t \frac{\dot{x}_{0}\delta\dot{x}_{0}-\ddot{x}_{0}\delta x_{0}}
{\dot{x}_{0}^{2}} dt' =\delta E_{0}\int_{\tn}^t \frac{1}{\dot{x}_{0}^{2}}dt',
\end{displaymath}
hence
\begin{displaymath}
\left[\frac{\delta \xn}{\dot{x}_{0}}\right]_{\tn}^t=\delta E_{0}\int_{\tn}^t
\frac{1}{\dot{x}_{0}^{2}}dt'
\end{displaymath}
and finally
\begin{equation}
\label{xijvar}
\frac{\delta \xn}{\delta E_{0}}=\dot{x}_{0} \int_{\tn}^t
\frac{1}{\dot{x}_{0}^{2}}dt',
\end{equation}
which is exactly the term with coefficient $E_{ij}$ in (\ref{solutxij}). 
So we end up with:
\begin{displaymath}
\xij=-c_{ij}\frac{\delta \xn}{\delta \tn} +E_{ij}\frac{\delta\xn}  
{\delta E_{0}} +\frac{\Rij}{2E_{0}}U.
\end{displaymath}

\bigskip
In the same way we can analyse the third order fermion solutions
(\ref{3rdvar1}) and (\ref{3rdvar2}). Both have parts proportional 
to $\dot{x}_{0}$ and $U$. But in addition there are two further terms 
which have to be interpreted as variations of these parts with 
$\tn$ and $E_{0}$. The $\tn$-variations can be written down immediately:
\begin{displaymath}
\frac{\delta \dot{x}_{0}}{\delta\tn} \propto \ddot{x}_{0},\;
\frac{\delta U}{\delta\tn} \propto U'\dot{x}_{0}
\end{displaymath}
and yield the terms with coefficients $c_{[ij}l_{k]}$ and
$c_{[ij}r_{k]}$. The $E_{0}$-variations give:
\begin{displaymath}
\frac{\delta \dot{x}_{0}}{\delta E_{0}}=\frac{d}{dt}\frac{\delta \xn}
{\delta E_{0}}=\ddot{x}_{0}\int_{\tn}^t
\frac{1}{\dot{x}_{0}^{2}}dt'+\frac{1}{\dot{x}_{0}},\; 
\frac{\delta U}{\delta E_{0}}=U'\frac{\delta \xn}{\delta E_{0}}=
U'\dot{x}_{0}\int_{\tn}^t \frac{1}{\dot{x}_{0}^{2}}dt'.
\end{displaymath}
They explain the first part of the terms with coefficients
$E_{[ij}l_{k]}$ and $E_{[ij}r_{k]}$; note that the second part can be absorbed
into the $l_{ijk}$ and $r_{ijk}$-terms.

So the vectors formed by $\lijk$ and $\rijk$ consist of three parts: A
vector rigidly fixed to the familiar first order fermionic vectors and
two parts which are proportional to the variation of these vectors with
$\tn$ and $E_{0}$. 

\bigskip
Finally we want to explain the parts of the highest order bosonic
function $x_{1234}$. The first three terms of (\ref{4var}) are
familiar: They are the variations of $\xn$ with $\tn$ and $E_{0}$ and
the inhomogeneity term proportional to $U$. The next three terms can
be explained as \textit{second} variations. So $\ddot{x}_{0}$ gives the
second variation of $\xn$ with respect to $\tn$; for the term with
coefficient $c_{[12}E_{34]}$ we have already shown that it is the time
derivative of $\frac{\delta \xn}{\delta E_{0}}$, hence it is
proportional to $\frac{\delta^2\!\xn}{\delta E_{0}\delta \tn}$. So we
only have to deal with the $E_{[12}E_{34]}$-component. We find that
\begin{displaymath}
\frac{\delta^2 \xn}{\delta E_{0}^{2}}=\frac{\delta}{\delta E_{0}}
\left(\dot{x}_{0}\int_{\tn}^t\frac{1}{\dot{x}_{0}^{2}}dt'\right)=
\frac{\delta \dot{x}_{0}}{\delta E_{0}}\int_{\tn}^t\frac{1}{\dot{x}_{0}^{2}}dt'
+\dot{x}_{0}\frac{\delta}{\delta E_{0}}
\int_{\tn}^t\frac{1}{\dot{x}_{0}^{2}}dt'.
\end{displaymath}
The last term can be rewritten as
\begin{eqnarray*}
\dot{x}_{0}\int_{\tn}^t\frac{-2}{\dot{x}_{0}^{3}}\frac{\delta
\dot{x}_{0}}{\delta E_{0}}dt'&=&-2 \dot{x}_{0}\int_{\tn}^t\left(
\frac{1}{\dot{x}_{0}^{4}}+\frac{\ddot{x}_{0}}{\dot{x}_{0}^{3}}
\int_{\tn}^t\frac{1}{\dot{x}_{0}^{2}}dt''\right)dt'\\&=&
-2\dot{x}_{0}\int_{\tn}^t\frac{1}{\dot{x}_{0}^{4}}dt'+\dot{x}_{0}
\left(\frac{1}{\dot{x}_{0}^{2}}\int_{\tn}^t\frac{1}{\dot{x}_{0}^{2}}dt'
-\int_{\tn}^t\frac{1}{\dot{x}_{0}^{4}}dt'\right).
\end{eqnarray*}
Combining this with the expanded first term
\begin{displaymath}
\ddot{x}_{0}\left(\int_{\tn}^t\frac{1}{\dot{x}_{0}^{2}}dt'\right)^{2}
+\frac{1}{\dot{x}_{0}}\int_{\tn}^t\frac{1}{\dot{x}_{0}^{2}}dt'
\end{displaymath}
we find exactly the component in (\ref{4var}) with coefficient
$E_{[12}E_{34]}$.

The last two terms in (\ref{4var}) are easy again: They are given by the two
first variations of the inhomogeneity term $\frac{\Rij}{2E_{0}}U$ with
$\tn$ and $E_{0}$, which we have already calculated above. Note that the term
$-\frac{U}{E_{0}}$ with coefficient $R_{[12}E_{34]}$ can be absorbed
into the $R_{1234}$ term.

\bigskip
In summary:
\begin{enumerate}
\item The lowest order bosonic function $\xn$ describes the
one-dimensional motion of a point particle with energy $E_{0}$ in the potential
$\frac{1}{2}U^2(\xn)$. It is completely unaffected by all higher order
bosonic and all fermionic functions and not influenced by any higher
order charges. There are two free parameters involved, an initial time
$\tn$ and the energy $E_{0}$.
\item The first order fermionic functions comprise two-dimensional
vectors $(\li,\ri)$. The lengths of these vectors and the angles
between them are fixed by charges arising from symmetry under
change of basis in the Grassmann algebra and from the second order
charge $R_{ij}$, and so are constant in time, i.e. all
fermionic vectors rigidly corotate. Their motion is in turn determined
by the bosonic vector $(\dot{x}_{0},U)$ which has constant squared length
$2E_{0}$ and is rigidly coupled to the fermion vectors by the supercharges
$Q_{i}$ and $\tilde{Q}_{i}$. For every generator there is a pair of
free parameters which specify the length and the phase of the
corresponding vector.
\item The second order bosonic function $\xij$ has terms proportional
to the (first) variation of $\xn$ with $\tn$ and $E_{0}$. There is one
additional term proportional to $U$ stemming from the inhomogeneity of
the $\xij$-equation of motion and ultimately from the Yukawa interaction term.
\item The third order fermionic functions again comprise
two-dimensional vectors $(\lambda_{ijk}, \rho_{ijk})$.  They can be
divided into three additive parts. The first part is rigidly rotating
with the first order fermionic vectors, the second and third parts are
proportional to the variations of this first order vectorial motion with 
initial time $\tn$ and $E_{0}$. 
\item Finally, the fourth order bosonic function involves all first
and second order variations of $\xn$ with respect to $\tn$ and $E_{0}$
plus a further term proportional to $U$ and \textit{its} 
first order variations.  
\end{enumerate}

\section{Oscillatory Motion}
\label{specialtypesofmotion}
In this section we apply our results of the previous section
to  oscillatory motion, i.e. we assume that the lowest order bosonic
function $\xn$ is periodic with period $T$. We have found already that
in this case the first order fermionic motion is described by
two-dimensional vectors that either continuously wind around a circle
or oscillate between two symmetrically lying points on it, depending
on the number of sign changes of $U$.

When we look at the next bosonic level, i.e. the
functions $x_{ij}$, we can see immediately that the first part of
the solution (\ref{solutxij}), which is proportional to $\dot{x}_{0}$, is a
periodic function with the same period $T$. The last term proportional
to $U(\xn(t))$ is periodic, too, but it can have period $\frac{1}{2}T$ if
the potential function $U(\xn)$ has reflection symmetry, i.e. if
$U(a+x)=U(a-x)$ for some constant $a$. Let therefore $\hat{T}$, which is either
$T$ or $\frac{1}{2}T$, denote the period of $U(x_{0}(t))$. The most
interesting term is the remaining second one containing the
integral. We will analyse this term from a mathematical point of view
using Floquet theory but before we do that we would like to understand
it from a more physical perspective. 

From equation (\ref{xijvar}) in the previous section we know that the 
integral term can be interpreted as the variation of the solution
$\xn$ with respect to the energy $E_{0}$. Because the variation is
infinitesimally small, we can make the assumption that the functional
form of the motion remains unchanged. We treat the motion as determined by
only two parameters, the period $T$ and a characteristic amplitude $A$, 
defined e.g. as the distance between the two turning points of the
motion. In the generic case $T$ and $A$ will be related by a
non-trivial function $T(A)$, therefore a change in energy means not
only a change in the amplitude $A$ but also in the period $T$, so:
\begin{displaymath}
\frac{dx_{0}}{dE_{0}}=\frac{dx_{0}}{dA}\frac{dA}{dE_{0}}=
\left(\frac{\partial x_{0}}{\partial A}+\frac{\partial x_{0}}{\partial T} 
\frac{dT}{dA}\right)\frac{dA}{dE_{0}}.
\end{displaymath}
Since $\xn$ is a $T$-periodic function we can use its Fourier series
$\sum_{j}c_{j} e^{i\omega_{j}t}$,
where $\omega_{j}=\frac{2\pi}{T}j$ and the coefficients $c_{j}$ are
regarded as functions of $A$, to find:
\begin{displaymath}
\frac{\partial x_{0}}{\partial T}=
-\frac{t}{T}\sum_{j} c_{j}i\omega_{j}e^{i\omega_{j}t}=
-\frac{\dot{x}_{0}t}{T}.
\end{displaymath}
Thus
\begin{displaymath}
\frac{dx_{0}}{dE_{0}}(t)=\left(\frac{\partial x_{0}}{\partial A}(t)
-t\frac{\dot{x}_{0}(t)}{T}\frac{dT}{dA}\right)\frac{dA}{dE_{0}}.
\end{displaymath}
Both $\frac{\partial x_{0}}{\partial A}$ and $\dot{x}_{0}$ are
$T$-periodic functions, but $\frac{dx_{0}}{dE_{0}}$ itself is clearly not
since the second term is \textit{linear}-periodic, i.e. the 
product of the linear term $t$ with a periodic function.
The extra linear-periodic term diverges with time, thus seemingly spoiling the
oscillatory nature of the solution. However, this problem only arises
because the period of the oscillation depends on the amplitude $A$ and
thus ultimately on the energy $E_{0}$. 
Expressing a solution with slightly changed period $T+dT$ in terms of
$T$-periodic functions inevitably leads to non-periodic terms. 
The effect is well-known from celestial mechanics where linear-periodic
functions appear as secular terms in the study of stability
problems of planetary orbits (see e.g. \cite{Goldstein}).

There is one case where no linear-periodic term occurs, and this is
when 
\begin{displaymath}
\frac{dT}{dA}=0,
\end{displaymath}
i.e. when the period is independent of the amplitude. This is true for
the harmonic oscillator, and consequently there is no non-periodic
term in equation (\ref{harmxijsolut}).

\bigskip
The existence of non-periodic terms in the bosonic function $x_{ij}$
can be derived in a mathematically more stringent way from Floquet
theory, the theory of linear differential equations with periodic
coefficients. Equation (\ref{xij}), with
$\lambda_{i}\rho_{j}-\rho_{i}\lambda_{j}$ replaced by the constant
$R_{ij}$ according to (\ref{Rij}), determines the motion of $x_{ij}$.
Because we know that the particular solution proportional to $U$ is
periodic it suffices to treat the homogeneous equation which can be
written in the form
\begin{equation}
\label{floquet}
\ddot{x}_{ij}+p(t)x_{ij}=0,
\end{equation}
where $p$ is a $\hat{T}$-periodic function. This allows us to apply
Floquet theory which states that \cite{Floquet}:
\begin{enumerate}
\item There exists a non-zero constant $\alpha$, called the characteristic
multiplier, and a non-trivial solution $x_{ij}(t)$ such that
\begin{equation}
\label{multiplier}
x_{ij}(t+\hat{T})=\alpha x_{ij}(t), 
\end{equation}
from which one deduces
\item There exist linearly independent solutions $x_{ij,1}$ and
$x_{ij,2}$ to (\ref{floquet}),
such that either
\begin{equation}
\label{floquet1}
x_{ij,1}(t)=e^{m_{1}t}P_{1}(t),\;x_{ij,2}(t)=e^{m_{2}t}P_{2}(t)
\end{equation}
or
\begin{equation}
\label{floquet2}
x_{ij,1}(t)=e^{m t}P_{1}(t),\;x_{ij,2}(t)=e^{mt}(tP_{1}(t)+P_{2}(t)),
\end{equation}
where in both cases $P_{1}, P_{2}$ are $\hat{T}$-periodic functions
and $m_{1}$, $m_{2}$, $m$, called characteristic exponents,  are --
not necessarily distinct -- constants. 
\end{enumerate}
Whether the solutions take the form (\ref{floquet1}) or
(\ref{floquet2}) depends on whether there are two independent
solutions of (\ref{floquet}) with the property (\ref{multiplier}) or
just one. If we denote by $X_{1}, X_{2}$ the two linearly independent solutions
of (\ref{floquet}) with  
\begin{displaymath}
X_{1}(0)=1, \dot{X}_{1}(0)=0, X_{2}(0)=0, \dot{X}_{2}(0)=1, (\tn=0\:\:\:
\mbox{for simplicity})
\end{displaymath}
and by $M(t)$ the matrix
\begin{displaymath}
\left(
\begin{array}{cc}
X_{1}(t)&X_{2}(t)\\
\dot{X}_{1}(t)&\dot{X}_{2}(t)
\end{array}
\right),
\end{displaymath}
we get a solution of type (\ref{floquet1}) if $M(\hat{T})$ is
diagonalizable and a solution of type (\ref{floquet2}) if $M(\hat{T})$
has Jordan normal form.

To determine the characteristic exponents $m_{i}$ it 
is useful to notice that our equation is not an arbitrary 
differential equation with periodic coefficients but an example of
Hill's equation which takes the form 
\begin{displaymath}
F(t)\ddot{x}_{ij}+F'(t)\dot{x}_{ij}+G(t)x_{ij}=0,
\end{displaymath}
$F$ and $G$ being $\hat{T}$-periodic functions. For equations of Hill's type
the Floquet multipliers are completely determined by the trace $D$ of the
fundamental matrix $M(\hat{T})$. This trace can be determined
indirectly, using the fact that we know one periodic solution \mbox{of
(\ref{floquet}),} namely $\dot{x}_{0}$, which has either period $\hat{T}$ or
$2\hat{T}$, depending on whether $U(x_{0})$ is reflection-symmetric or not. Now
we can use another theorem of Floquet theory which states:
\begin{enumerate}
\item Hill's equation has non-trivial solutions with period $\hat{T}$
if and only if $D=2$. Then either $m_{1}=m_{2}=m=0$ and
\begin{displaymath}
x_{ij,1}(t)=P_{1}(t),\;x_{ij,2}(t)=P_{2}(t),
\end{displaymath}
or $m=0$ and
\begin{displaymath}
x_{ij,1}(t)=P_{1}(t),\;x_{ij,2}(t)=tP_{1}(t)+P_{2}(t).
\end{displaymath}
\item Hill's equation has non-trivial solutions with period $2\hat{T}$
if and only if $D=-2$. Then either $m_{1}=m_{2}=\frac{i\pi}{\hat{T}}$ and
\begin{displaymath}
x_{ij,1}(t)=e^{\frac{i\pi}{\hat{T}}t}P_{1}(t),\;
x_{ij,2}(t)=e^{\frac{i\pi}{\hat{T}}t}P_{2}(t),
\end{displaymath}
or $m=\frac{i\pi}{\hat{T}}$ and
\begin{displaymath}
x_{ij,1}(t)=e^{\frac{i\pi}{\hat{T}}t}P_{1}(t),\;
x_{ij,2}(t)=e^{\frac{i\pi}{\hat{T}}t}(tP_{1}(t)+P_{2}(t)).
\end{displaymath}
\end{enumerate}
Using these theorems we find that the second independent solution
(\ref{xijvar}) has to be either periodic or linear-periodic. The
period is in both cases $T$, the period of the first independent
solution $\dot{x}_{0}$. This means that we can formally confirm our
earlier result derived from physical arguments. Additionally, we have
that any solution must be bounded in finite intervals. This is not
immediately obvious from the explicit form of (\ref{xijvar}), which
could be singular when $\dot{x}_{0}=0$.

\bigskip
Before we proceed to analyse a particular potential function which
admits oscillatory motion we want to comment briefly on the higher
order bosonic and fermionic quantities $\lambda_{ijk}, \rho_{ijk}$ and
$x_{1234}$. From equations (\ref{3rdvar1}) and (\ref{3rdvar2}) we can
see that all terms that appear in the third order fermionic solution
are either periodic or linear-periodic. For the term
\begin{displaymath}
\ddot{x}_{0}\int_{\tn}^{t}\frac{1}{\dot{x}_{0}^{2}}dt'+\frac{1}{\dot{x}_{0}}
\end{displaymath}
this follows e.g. from the fact that the derivative of a linear-periodic
term is again linear-periodic. We can expect this behaviour physically
since $\lambda_{ijk}$ and $\rho_{ijk}$ are given by the \textit{first}
variation of the motion of the first order fermionic vectors with
energy and initial time. (There are further periodic contributions to
$\lambda_{ijk}$ and $\rho_{ijk}$ with coefficients $l_{ijk}$ and $r_{ijk}$.)
 
Most terms in the fourth order bosonic solution $x_{1234}$ can be
easily seen to be periodic or linear-periodic in the same
manner. However, there are the three terms which have been interpreted
as the \textit{second} variation of $\xn$ with energy $E_{0}$. These
should contain a quadratic term in $t$, because
\begin{displaymath}
\frac{d^2\xn}{dE_{0}^2}=\left[\frac{\partial^{2}\xn}{\partial
A^{2}}+2\frac{\partial^{2}\xn}{\partial A \partial T}\frac{dT}{dA}+
\frac{\partial^{2}\xn}{\partial T^{2}} \left(\frac{dT}{dA}\right)^{2} 
+\frac{\partial \xn}{\partial T}\frac{d^{2}T}{dA^{2}} \right]
\left(\frac{dA}{dE_{0}}\right)^{2}+\left[\frac{\partial \xn}{\partial A}
+\frac{\partial \xn}{\partial T}\frac{dT}{dA}\right] \frac{d^{2}A}{dE_{0}^{2}}
\end{displaymath}
and
\begin{displaymath}
\frac{\partial^{2}\xn}{\partial T^{2}}=\frac{\partial}{\partial
T}\left(-\frac{\dot{x}_{0} t}{T}\right)=
\frac{1}{T^{2}}\left(2\dot{x}_{0}t+\ddot{x}_{0}t^{2}\right),
\end{displaymath}
all other terms being (linear-)periodic. Again, the quadratic
term does not occur when period $T$ and amplitude $A$ are independent,
i.e. when $\frac{dT}{dA}=0$. This is confirmed by the harmonic
oscillator example where $x_{1234}$ is completely periodic. 

We can also understand the existence of a quadratic-periodic term directly
from the \mbox{solution (\ref{4var})} which contains the term
\begin{displaymath}
\ddot{x}_{0}\left(\int_{\tn}^{t}\frac{1}{\dot{x}_{0}}dt'\right)^{2}.
\end{displaymath} 
We know already that the integral will yield a term proportional to
$t$; this is squared to give a quadratic term in $t$, and then
multiplied by the periodic function $\ddot{x}_{0}$. This quadratic
term cannot cancel out because the two remaining terms with
coefficient $E_{[12}E_{34]}$ in (\ref{4var}) are at most linear-periodic.

\subsection{The hyperbolic potential $U(x)=c \tanh kx$}
\label{hyperbolpot}
We illustrate our results by the hyperbolic potential $U(x)=c \tanh
kx$, already discussed in section \ref{grassmannhyperbol} for general
Grassmann algebra. Since we are only interested in oscillatory motion
here, we make the assumption that $c^{2}>2E_{0}$.

Inverting the result obtained from (\ref{solutxn}) we can calculate that 
\begin{samepage}
\begin{equation}
\xn(t)=\pm\frac{1}{k} \mbox{arcsinh}
\left(
\begin{array}{l}
\sqrt{\frac{2E_{0}}{c^{2}-2E_{0}}}
\end{array}
\sin (\omega(t\!-\!\tn)+\kappa)\right), 
\label{xnhyperbol}
\end{equation}
with
\begin{displaymath}
\begin{array}{lcl}
\omega=k\sqrt{c^{2}-2E_{0}}&,&
\kappa=\arcsin\left(\sqrt{\frac{c^{2}}{2E_{0}}-1}
\sinh k x_{0}(\tn)\right).
\end{array}
\end{displaymath}
So $\xn$ is indeed a periodic function.
\end{samepage}
For $\dot{x}_{0}$ and $U$ we immediately find that
\begin{eqnarray}
\dot{x}_{0}(t)&=&\pm \frac{\sqrt{c^{2}-2E_{0}}}{\sqrt{\frac{c^{2}}{2E_{0}}
\sec^{2}y -1}}\label{hyperbolxdot}\\
U(t)&=&\pm\frac{c\tan y}
{\sqrt{\frac{c^{2}}{2E_{0}}\sec^{2}y -1}},\label{hyperbolU} 
\end{eqnarray}
where we have abbreviated $(\omega(t\!-\!\tn)+\kappa)$ to $y$.
Therefore the motion of the bosonic \mbox{vector (\ref{bosvector})} is a
\textit{non-uniform rotation}, which confirms our comments in
section \ref{interpret} since the potential function $U$ changes sign
only once. All fermionic vectors corotate with the bosonic one.

To determine $x_{ij}$ we need the integral
\begin{displaymath}
\int_{\tn}^{t}\frac{1}{\dot{x}_{0}^{2}}dt'= 
\frac{1}{k\sqrt{(c^{2}\!-\!2E_{0})^{3}}} 
\left(-\omega(t\!-\!\tn)+\frac{c^{2}}{2E_{0}} \tan y\right);
\end{displaymath}
we then obtain for the $E_{ij}$-related part of $x_{ij}$:
\begin{equation}
\pm\frac{E_{ij}}{k(c^{2}\!-\!2E_{0})}
\left(\frac{\frac{c^{2}}{2E_{0}}\tan y}
{\sqrt{\frac{c^{2}}{2E_{0}}\sec^{2}y\!-\!1}}
-\frac{\omega(t\!-\!\tn)}{\sqrt{\frac{c^{2}}{2E_{0}}\sec^{2}y\!-\!1}}
\right).
\label{hyperbolxij}
\end{equation}
This result clearly shows a linear periodic term which means
that the period-amplitude relation is non-trivial. In fact, period and
amplitude of $x_{0}(t)$, the latter defined as distance between the
two turning points, can be easily calculated using (\ref{xnhyperbol}):
\begin{displaymath}
T=\frac{2\pi}{k\sqrt{c^{2}-2E_{0}}},\;
A=\frac{2}{k}\mbox{arcsinh} \sqrt{\frac{2E_{0}}{c^{2}-2E_{0}}}.
\end{displaymath}
Thus
\begin{displaymath} 
T(A)=\frac{2\pi}{k|c|}\cosh{\frac{kA}{2}}
\end{displaymath}
and $\frac{dT}{dA}\neq 0$. We point out that all terms in the solution
are clearly bounded, which is not the case for the integral
itself.

\begin{figure}[ht]
\begin{center}
\epsfig{file=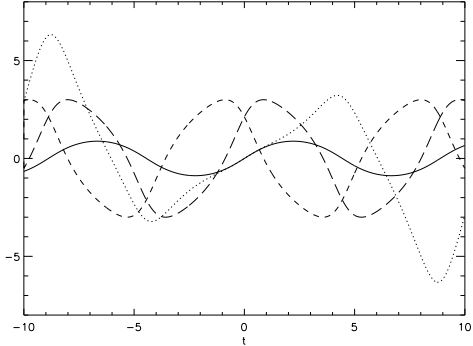,width=7cm}
\caption{{\small \it $x_{0}$ (continuous), $x_{ij}$ (dotted), $\lambda_{i}$
(long dashes) and $\rho_{i}$ (short dashes) as functions of time for
$c=k=1$, $x_{0}(0)=c_{ij}=0$, $E_{0}=\frac{1}{4}$,
$E_{ij}=\frac{1}{2}$ and $Q_{i}=\tilde{Q}_{i}=\frac{3}{2}$}}
\label{diagram2}
\end{center}
\end{figure}

Figure \ref{diagram2} gives an example of both the bosonic quantities
$x_{0}$ and  $x_{ij}$ and the fermionic quantities $\lambda_{i}$ and
$\rho_{i}$. 

We do not present the higher order solutions in detail here since their
explicit form, although not complicated to find, is lengthy and does
not provide new insights. The only interesting exception is the part
of $x_{1234}$ which has been characterized as a second variation with energy.
This part can be calculated to be
\begin{eqnarray*}
&&\frac{1}{2}E_{[12}E_{34]}\left[
\frac{(\alpha^{3}\!-\!3\alpha^{2}\!+\!6\alpha)\sin y\sec^{3}y
+(2\alpha^{2}\!-\!4\alpha)\sin y \sec y}
{k(c^{2}\!-\!2E_{0})^{2}\sqrt{(\alpha \sec^{2}y\!-\!1)^{3}}}\right.\\
&&\left.+(t-\tn)\left(\frac{\left(-2\alpha^{2}
+\alpha\right)\sec^{2}y+1}{\sqrt{(c^{2}\!-\!2E_{0})^{3}}
\sqrt{(\alpha\sec^{2}y\!-\!1)^{3}}}\right)
+(t-\tn)^{2}\left(-\frac{k\alpha}{c^{2}\!-\!2E_{0}}
\frac{\sin y \sec^{3}y}{\sqrt{(\alpha \sec^{2}y\!-\!1)^{3}}}
\right)
\right],
\end{eqnarray*}
where we have abbreviated $\frac{c^{2}}{2E_{0}}$ to $\alpha$. Again,
there are two noteworthy points here: Firstly, the occurence of a
quadratic-periodic term, which has been predicted in section
\ref{interpret} and secondly the complete boundedness of the whole
solution, which is not immediately obvious from the functional form
(\ref{4var}), especially regarding the term where the velocity appears
in the denominator.

\bigskip
Next we want to show that the two solution methods presented in this
paper deliver the same result. Therefore we have to decompose our
previous solution, the first of equations (\ref{xnhyperbolgrassmann}),
from section \ref{grassmannhyperbol}. Recall that any real analytical
function $f$ is extended to a Grassmann-valued function by the formula
\begin{displaymath}
f(z)=\sum_{n=0}^{\infty}\frac{1}{n!}f^{n}(z_{b})z_{s}^{n},
\end{displaymath}
where $z_{b}$ and $z_{s}$ are body and soul of the Grassmann
number $z$, respectively \cite{deWitt}. To keep the calculation short we will expand
here only up to second order, i.e. the following equations have to be
understood modulo a component proportional to $\xi_{1234}$.

We begin with the frequency $\omega$:
\begin{displaymath}
\omega=k\sqrt{(c^{2}\!-\!2E_{0})\!-\!2E_{ij}\xi_{ij}}
=k\sqrt{c^{2}\!-\!2E_{0}}-\frac{kE_{ij}}{\sqrt{c^{2}\!-\!2E_{0}}}\xi_{ij}.
\end{displaymath}
Then
\begin{eqnarray*}
\sin(\omega(t\!-\!\tn)+\kappa)&=&
\sin(k\sqrt{c^{2}\!-\!2E_{0}}(t\!-\!\tn)\!+\!\kappa_{0})\\
&&+\cos(k\sqrt{c^{2}\!-\!2E_{0}}(t\!-\!\tn)\!+\!\kappa_{0})
\left(-\frac{kE_{ij}(t\!-\!\tn)}{\sqrt{c^{2}\!-\!2E_{0}}}
\!+\!\kappa_{a,ij}\right)\xi_{ij}.
\end{eqnarray*}
Since
\begin{displaymath}
\sqrt{\frac{2E\!+\!2ckR}{c^{2}\!-\!2E}}=\sqrt{\frac{2E_{0}}{c^{2}\!-\!2E_{0}}}
+\left(\frac{c^{2}E_{ij}+ckR_{ij}(c^{2}-2E_{0})}
{\sqrt{2E_{0}(c^{2}\!-\!2E_{0})^{3}}}\right)\xi_{ij}
\end{displaymath}
we find
\begin{eqnarray*}
\sqrt{\frac{2E\!+\!2ckR}{c^{2}\!-\!2E}}
\sin(\omega(t\!-\!\tn)\!+\!\kappa)
&=&\sqrt{\frac{2E_{0}}{c^{2}\!-\!2E_{0}}}\sin y
+\sqrt{\frac{2E_{0}}{c^{2}\!-\!2E_{0}}}
\left[\frac{R_{ij}}{2E_{0}}ck\sin y
+\kappa_{ij}\cos y\right.\\
&&\left.+E_{ij}\left(\frac{\frac{c^{2}}{2E_{0}}}
{c^{2}\!-\!2E_{0}}
\sin y -\frac{k(t\!-\!\tn)}{\sqrt{c^{2}\!-\!2E_{0}}}\cos y
\right)\right]\xi_{ij},
\end{eqnarray*}
where again $y=k\sqrt{c^{2}-2E_{0}}(t\!-\!\tn)+\kappa_{0}$. This
result equals $\sinh kx$; decomposition yields:
\begin{displaymath}
\sinh kx=\sinh(k\xn+kx_{ij}\xi_{ij})=\sinh kx_{0}+kx_{ij}(\cosh kx_{0})\xi_{ij},
\end{displaymath}
thus by comparison
\begin{displaymath}
\sinh kx_{0}=\sqrt{\frac{2E_{0}}{c^{2}\!-\!2E_{0}}}\sin y,
\end{displaymath}
which gives the same solution for $\xn$ as (\ref{xnhyperbol}). 
Similarly, we can read off the result for $kx_{ij}\cosh kx_{0}$ and
then calculate $x_{ij}$:
\begin{eqnarray*}
x_{ij}(t)&=&\frac{R_{ij}}{2E_{0}}\frac{c\tan y}
{\sqrt{\frac{c^{2}}{2E_{0}}\sec^{2}y-1}}-\frac{\kappa_{ij}}{k}\frac{1}
{\sqrt{\frac{c^{2}}{2E_{0}}\sec^{2}y-1}}\\
&&+E_{ij}\left(\frac{\frac{c^{2}}{2E_{0}}}{k(c^{2}\!-\!2E_{0})}\frac{\tan y}
{\sqrt{\frac{c^{2}}{2E_{0}}\sec^{2}y-1}}-\frac{1}{\sqrt{c^{2}\!-\!2E_{0}}}
\frac{t-\tn}{\sqrt{\frac{c^{2}}{2E_{0}}\sec^{2}y-1}}\right).
\end{eqnarray*}
The term with coefficient $R_{ij}$ can be readily identified as the
potential function $U(t)$, derived in (\ref{hyperbolU}). The second term with
coefficient $\kappa_{ij}$ equals the velocity $\dot{x}_{0}$, which
we have calculated in (\ref{hyperbolxdot}), when we make the identification
\mbox{$\kappa_{ij}=-\omega c_{ij}$}. Finally, the term
proportional to $E_{ij}$ is identical with our \mbox{result
(\ref{hyperbolxij})}.

So the solution found by using our layer-by-layer approach can also be
obtained by the decomposition of the full Grassmann 
solution calculated through direct integration in Grassmann space. We
have demonstrated this feature here using the bosonic quantity $x(t)$
but we could equally well have chosen one of the fermionic quantities
$\psi_{+}(t)$ or $\psi_{-}(t)$. This means that our two solution
methods are compatible with each other.

\section{The Zero Energy Solutions}
\label{zeroenergy}
We have already mentioned in sections \ref{solfixpot} and \ref{solarbpot}
that the case where the energy $E_{0}$, the body of the full Grassmann
energy $E$, equals zero has to be treated differently.
All the solutions to the equations of motion (\ref{x0})--(\ref{lijkrijk}) were
given under the restriction $E_{0} \neq 0$ so that we cannot use our
previous results. It will turn out, however, that for a finitely
generated Grassmann algebra we can derive all solutions in explicit
functional form for arbitrary potential.

\bigskip
Starting with equation (\ref{H0}) we can see that if $E_{0}=0$ both
$\dot{x}_{0}$ and $U(\xn)$ have to vanish:
\begin{displaymath}
\dot{x}_{0}=0,\:U(\xn)=0.
\end{displaymath}
This means that the particle stays permanently at rest at a minimum
$x_{0,min}$ of the \mbox{potential $U^{2}$.} One could assume at this
point that all higher order bosonic and fermionic functions are
trivial as well, but as we shall soon see this is far from being
necessary, in contradiction to \cite{Junker}.  

Because $\xn(t)$ is constant, all the (spatial) derivatives of $U$
are constant functions of time too; we denote their values by 
\begin{displaymath}
U'(\xn)= k_{1},\,U''(\xn)= k_{2},\,U'''(\xn)= k_{3}, \dots;
\end{displaymath}
we assume in the following that $k_{1}\neq 0$.

\bigskip
When we now look at the first order fermionic equations (\ref{liri})
for $\lambda_{i}$ and $\rho_{i}$, we find that they are easily solved by
the harmonic oscillator ansatz:
\begin{eqnarray}
\lambda_{i}&=&l_{i}\cos k_{1}(t-\tn) - r_{i} \sin k_{1}(t-\tn)\label{E0li}\\
\rho_{i}&=&l_{i}\sin k_{1}(t-\tn) + r_{i} \cos k_{1}(t-\tn).\label{E0ri}
\end{eqnarray}
Unlike the true harmonic oscillator, the constants $l_{i}$ and
$r_{i}$ are, however, not linked to the first order supercharges
$Q_{i}$ and $\tilde{Q}_{i}$ which vanish completely as can be verified
from (\ref{qiqti}).

To understand the first order fermionic motion we assume that the
energy $E_{0}$ is small but non-zero, restricting the particle motion
to a small neighbourhood of the stability point $x_{0,min}$ where the
potential function $U$ is approximately linear. The result is an
almost harmonic oscillation with frequency $k_{1}$, mirrored
by the fermionic quantities -- as follows from (\ref{fermionharm1}) and
(\ref{fermionharm2}). When we now let $E_{0}$ approach zero, this
fermionic motion seems to diverge as one can see from the formulae for
the coefficients $L_{i}$ and $R_{i}$ in (\ref{Li}) and (\ref{Ri}). 
However, there is a subtle point here: While the supercharges $Q_{i}$
and $\tilde{Q}_{i}$ are arbitrarily chosen constants for $E_{0}\neq0$,
they have to vanish for $E_{0}=0$. Thus to avoid any discontinuities 
they have to smoothly approach zero as the energy decreases. Exactly 
how they approach zero finally determines which value the harmonic
oscillator coefficients $L_{i}$ and $R_{i}$ take in the limit $E_{0}=0$, 
which justifies the two remaining degrees of freedom in our
solution, $l_{i}$ and $r_{i}$. 

\bigskip
Next we want to analyse the second order bosonic quantity $x_{ij}$.
Because $U'$ and $U''$ are constants the equation of motion
(\ref{xij}) describes formally a harmonic oscillator with frequency
$k_{1}$ subject to a constant external force $k_{2}R_{ij}$,
so the solution (for $k_{1} \neq 0$) is
\begin{equation}
\label{E0xij}
x_{ij}=v_{ij}\cos k_{1}(t-\tn) + \tilde{v}_{ij}\sin k_{1}(t-\tn)
+\frac{k_{2}}{k_{1}^{2}}R_{ij},
\end{equation}
where $v_{ij}$ and $\tilde{v}_{ij}$ are integration constants. To
interpret this result we recall that in case $E_{0}\neq0$ the
homogeneous part of $x_{ij}$ consists of two terms, the two
variational derivatives of $\xn$ with respect to $E_{0}$ and $\tn$. When  
$E_{0}$ is zero, a small change in energy will result in oscillatory
motion around the stability point $x_{0,min}$. This explains the
functional form and one of the free parameters of the solution
(\ref{E0xij}). The second parameter, though, cannot be connected to
the variation with $\tn$ anymore since this variation is zero for a
constant $\xn$. 

An interesting observation is connected to the second order
energy which can be calculated from (\ref{Hij}) to be 
$E_{ij}=-k_{1}R_{ij}$. Thus the second order energy is not independent
any more but determined by the four first order constants $l_{i}$,
$r_{i}$, $l_{j}$, $r_{j}$. Note the fact that $E_{ij}$ is \textit{not}
connected to the parameters $v_{ij}$ and $\tilde{v}_{ij}$.

\bigskip
The third order fermion equations (\ref{lijkrijk}) are again equations
describing a harmonic oscillator with a driving term. The homogeneous
part of the solution looks therefore the same as (\ref{E0li}) and
(\ref{E0ri}) with integration constants $l_{ijk}$ and $r_{ijk}$. To
find the particular solution we have to investigate the driving terms
$-k_{2} x_{[ij}\rho_{k]}$ and $k_{2} x_{[ij}\lambda_{k]}$
further. Using equations (\ref{E0li}), (\ref{E0ri}) and (\ref{E0xij}) we find
\begin{eqnarray*}
x_{[ij}\rho_{k]}&=&C_{ijk}+D_{1,ijk}\sin 2k_{1}(t-\tn) 
- D_{2,ijk} \cos 2k_{1} (t-\tn)\\
x_{[ij}\lambda_{k]}&=&\tilde{C}_{ijk}+D_{2,ijk} \sin 2k_{1}(t-\tn)
+D_{1,ijk} \cos 2k_{1}(t-\tn),
\end{eqnarray*}
where $C_{ijk}$, $\tilde{C}_{ijk}$, $D_{1,ijk}$ and $D_{2,ijk}$ are
constants built from $l_{i}$, $r_{i}$, $v_{ij}$
and $\tilde{v}_{ij}$. So the driving terms are not constant but have
an oscillating part that oscillates with \textit{double} the basic
frequency $k_{1}$. The particular solutions are thus
\begin{eqnarray*}
\lambda_{ijk,part.}&=&-\frac{k_{2}}{k_{1}}\tilde{C}_{ijk}+
\frac{k_{2}}{k_{1}}D_{1,ijk}\cos 2k_{1}(t-\tn)+
\frac{k_{2}}{k_{1}}D_{2,ijk}\sin 2k_{1}(t-\tn)\\
\rho_{ijk,part.}&=&-\frac{k_{2}}{k_{1}}C_{ijk}+
\frac{k_{2}}{k_{1}}D_{1,ijk}\sin 2k_{1}(t-\tn)-
\frac{k_{2}}{k_{1}}D_{2,ijk}\cos 2k_{1}(t-\tn).
\end{eqnarray*}
The fermionic vectors $(\lambda_{ijk}, \rho_{ijk})$ consist of three
parts: A uniform rotation with natural frequency $k_{1}$, a forced
uniform rotation with double this frequency that we can interpret as the
\textit{first excited} mode and a constant shift away from the origin. 
Note that there is no double frequency term if $U''(x_{0})=0$, and in
particular for the harmonic oscillator potential. This agrees with our
earlier result in section \ref{harmpot}.  
 
Calculating the supercharges $Q_{ijk}$ and $\tilde{Q}_{ijk}$ from
(\ref{qiqti}) we find that $Q_{ijk}=2k_{1}C_{ijk}$ and
$\tilde{Q}_{ijk}=-2k_{1}\tilde{C}_{ijk}$ which gives us a neat
interpretation of the displacement constants $C_{ijk}$ and
$\tilde{C}_{ijk}$. It further means that the third order supercharges
are not related to the free parameters $l_{ijk}$ and $r_{ijk}$ as is
the case for $E_{0}\neq 0$ but instead are completely 
determined by first and second order parameters. This is the same
phenomenon already encountered for the second order energy $E_{ij}$
and again it means that we should not think of $Q_{ijk}$ and
$\tilde{Q}_{ijk}$ as independent constants when $E_{0}=0$.
 
\bigskip
Finally we have to investigate the fourth order bosonic quantity
$x_{1234}$. The equation of motion (\ref{x1234}) simplifies in the
case $E_{0}=0$ to
\begin{displaymath}
\ddot{x}_{1234}+k_{1}^2x_{1234}= -\frac{3}{2}k_{1}k_{2}x_{[12}x_{34]}+
k_{2}R_{1234}+k_{3}x_{[12}R_{34]},
\end{displaymath}
so it describes a harmonic oscillator with three driving terms. The
homogeneous part is analogous to (\ref{E0xij}) with two integration
constants $c_{1234}$ and $\tilde{c}_{1234}$; the comments made about
$x_{ij}$ apply here, too. Analyzing the three driving terms we first
note that $k_{2}R_{1234}$ specifies a constant external force, playing
the same role as $k_{2}R_{ij}$ for the variables $x_{ij}$. The first
and third terms can be calculated using the explicit formulae for
$x_{ij}$, resulting in 
\begin{eqnarray*}
-\frac{3}{2}k_{1}k_{2}x_{[12}x_{34]}+k_{3}x_{[12}R_{34]}
&=&-\frac{3}{4}k_{1}k_{2}\left(C_{1234}+D_{1234}
\cos2k_{1}(t\!-\!\tn)+\tilde{D}_{1234}\sin 2k_{1}(t\!-\!\tn)\right)\\  
&&+\left(k_{3}\!-\!3\frac{k_{2}^{2}}{k_{1}}\right)\left(F_{1234} 
\cos k_{1}(t\!-\!\tn) +\tilde{F}_{1234} \sin k_{1}(t\!-\!\tn)\right),
\end{eqnarray*}
where $C_{1234}$, $D_{1234}$, $\tilde{D}_{1234}$, $F_{1234}$ and
$\tilde{F}_{1234}$ are fixed constants built from $v_{ij}$,
$\tilde{v}_{ij}$ and $R_{ij}$. So besides a further constant term we find
four oscillating ones, two with frequency $k_{1}$ and two with
frequency $2k_{1}$. The latter ones will cause forced harmonic motion 
with double the natural frequency as we found for the third order
fermionic terms. New, however, are the oscillating terms which have
the \textit{same} frequency as the homogenous solution. This means that the
oscillator is driven in resonance. Therefore the solution is not
bounded anymore but includes linear-periodic terms like $t \sin
k_{1}(t\!-\!\tn)$ and $t \cos k_{1}(t\!-\!\tn)$. Combining all contributions we
find the particular solution 
\begin{eqnarray*}
x_{1234,part.}&=&\frac{k_{2}}{k_{1}}
\left(\frac{R_{1234}}{k_{1}}\!-\!\frac{3}{4}C_{1234}\right)
+\frac{1}{4}\frac{k_{2}}{k_{1}}\left(D_{1234}\cos 2k_{1}(t\!-\!\tn)+
\tilde{D}_{1234}\sin 2k_{1}(t\!-\!\tn)\right)\\
&&+\frac{1}{2}\left(3\frac{k_{2}^{2}}{k_{1}^{2}}\!-\!\frac{k_{3}}{k_{1}}\right)
\left(\tilde{F}_{1234}\, t\cos k_{1}(t\!-\!\tn) - F_{1234}\, t\sin
k_{1}(t\!-\!\tn)\right).
\end{eqnarray*}
Thus the fourth order bosonic solution consists of harmonic
oscillation with the natural frequency $k_{1}$, an excited mode with
twice this frequency, linear-periodic motion and a constant shift.

Especially interesting are the linear-periodic terms. They can be
explained physically as in section \ref{specialtypesofmotion}, although
we have to see them as the \textit{second} variation with energy
here: When the particle stays at rest in a potential minimum the first
variation gives rise to harmonic oscillation, independently of the shape of
the potential function $U$ (as long as $k_{1}\neq 0$). Only when we
look at the second order variation do non-harmonic terms come into
play: The period of oscillation will in general depend on the
amplitude and the energy, therefore a variation in energy results in a
change of period; expressing the new solution in terms of the old
period then generally leads to secular, non-periodic terms.

Notice the fascinating fact that although the physical interpretation is
similar, the way the linear-periodic terms arise mathematically is
completely different: For non-zero energy oscillations they come as
solutions of a \textit{homogenous} differential equation with \textit{periodic}
coefficients; in the zero energy case they are particular solutions to
an \textit{inhomogenous} differential equation with \textit{constant}
coefficients. 

\section{Discussion}
In this work we have analysed a supersymmetric mechanical model from
two different viewpoints: Either we make no specification with regard
to the nature of the underlying Grassmann algebra $\mathcal{B}$. Then
for a range of potential functions $U$ the model can be explicitly
solved. Or we regard $\mathcal{B}$ as finitely generated thus reducing
all quantities to a set of real functions and their interrelationships.
Then we are able to solve the system completely, without any restrictions
on $U$. The methods have been shown to be compatible with each other here.

An open question with our first point of view is what meaning we can
give to the full, i.e. non-decomposed Grassmann solutions $x(t),
\psi_{+}(t)$ and $\psi_{-}(t)$. Interpretation may be aided by our
second approach, where we have shown that the component solutions
include the purely classical motion dependent on $E_{0}$ \textit{and}
all its variations with respect to this (real) constant.
Apparently, supersymmetric dynamics captures information about a whole
range of energies of the mechanical system, so in a sense we can say
that the Grassmann energy $E$ corresponds to some fuzzy classical
energy. This suggests possibly that supersymmetric classical dynamics
is closely related to the quantum dynamics. A study of the quantized
version of our model would therefore be an important step to
complement this research. 

We also hope to apply our methods to the full, i.e. space- and
time-dependent field theory.  

\section*{Acknowledgements}
N.S.M. thanks P. Bongaarts and R. Casalbuoni for correspondence, and
A. Rogers for discussions. R.H. gratefully acknowledges the support of
EPSRC and the Studienstiftung des Deutschen Volkes.

\end{document}